\documentclass[twocolumn]{aastex61}

\newcommand{\gy}{\textsc{Gyoto}\xspace}
\newcommand{\lo}{\textsc{Lorene}\xspace}
\newcommand{\lonr}{\textsc{Lorene/nrotstar}\xspace}
\newcommand{\at}{\textsc{Atm24}\xspace}
\newcommand{\be}{\begin{equation}}
\newcommand{\ee}{\end{equation}}
\newcommand{\bea}{\begin{eqnarray}}
\newcommand{\eea}{\end{eqnarray}}
\newcommand{\nn}{\nonumber}
\newcommand{\pp}{\varphi}
\newcommand{\dd}{\mathrm{d}}

\usepackage{amsmath}
\usepackage{xspace}

\received{...}
\revised{...}
\accepted{...}

\submitjournal{ApJ}

\shorttitle{Neutron star spectra from realistic atmospheres}
\shortauthors{Vincent et al.}

\begin{document}

 \title{Accurate ray-tracing of realistic neutron star atmospheres \\
   for constraining their parameters}
\correspondingauthor{Frederic H. Vincent}
\email{frederic.vincent@obspm.fr}

\author{Frederic H. Vincent}
\affil{LESIA, Observatoire de Paris, PSL Research University, CNRS, Sorbonne Universit\'es, UPMC Univ. Paris 06, Univ. Paris Diderot, Sorbonne Paris Cit\'e,
5 place Jules Janssen, 92195 Meudon, France}

\author{Micha\l \xspace Bejger}
\affil{Nicolaus Copernicus Astronomical Center, Polish Academy of Sciences,
Bartycka 18, PL-00-716 Warszawa, Poland}

\author{Agata R\'o\.za\'nska}
\affil{Nicolaus Copernicus Astronomical Center, Polish Academy of Sciences,
Bartycka 18, PL-00-716 Warszawa, Poland}

\author{Odele Straub}
\affil{LESIA, Observatoire de Paris, PSL Research University, CNRS, Sorbonne Universit\'es, UPMC Univ. Paris 06, Univ. Paris Diderot, Sorbonne Paris Cit\'e,
5 place Jules Janssen, 92195 Meudon, France}
\affil{LUTh, Observatoire de Paris, PSL Research University, CNRS UMR 8109,
Universit\'e Pierre et Marie Curie, Universit\'e Paris Diderot,
5 place Jules Janssen, 92195 Meudon, France}

\author{Thibaut Paumard}
\affil{LESIA, Observatoire de Paris, PSL Research University, CNRS, Sorbonne Universit\'es, UPMC Univ. Paris 06, Univ. Paris Diderot, Sorbonne Paris Cit\'e,
5 place Jules Janssen, 92195 Meudon, France}

\author{Morgane Fortin}
\affil{Nicolaus Copernicus Astronomical Center, Polish Academy of Sciences,
Bartycka 18, PL-00-716 Warszawa, Poland}

\author{Jerzy Madej}
\affil{Astronomical Observatory, University of Warsaw, Al. Ujazdowskie 4, PL-00-478 Warszawa, Poland}

\author{Agnieszka Majczyna}
\affil{National Centre for Nuclear Research, ul. Andrzeja So{\l }tana 7, 05-400 Otwock, Poland}

\author{Eric Gourgoulhon}
\affil{LUTh, Observatoire de Paris, PSL Research University, CNRS UMR 8109,
Universit\'e Pierre et Marie Curie, Universit\'e Paris Diderot,
5 place Jules Janssen, 92195 Meudon, France}

\author{Pawe\l \xspace Haensel}
\affil{Nicolaus Copernicus Astronomical Center, Polish Academy of Sciences,
Bartycka 18, PL-00-716 Warszawa, Poland}

\author{Leszek Zdunik}
\affil{Nicolaus Copernicus Astronomical Center, Polish Academy of Sciences,
Bartycka 18, PL-00-716 Warszawa, Poland}

\author{Bartosz Beldycki}
\affil{Nicolaus Copernicus Astronomical Center, Polish Academy of Sciences,
Bartycka 18, PL-00-716 Warszawa, Poland}


\begin{abstract}
    Thermal dominated X-ray spectra of neutron stars in quiescent transient X-ray binaries and neutron stars that undergo thermonuclear bursts are sensitive to mass and radius.   The mass-radius relation of neutron stars depends on the equation of state that governs their interior.   Constraining this relation accurately is thus of fundamental importance to understand the nature of dense matter.   In this context we introduce a pipeline to calculate realistic model spectra of rotating neutron stars with hydrogen and helium atmospheres.   An arbitrarily fast rotating neutron star with a given equation of state generates the spacetime in which the atmosphere emits radiation.   We use the \lonr code to compute the spacetime numerically and the \at code to solve the radiative transfer equations self-consistently.   Emerging specific intensity spectra are then ray-traced through the neutron star's spacetime from the atmosphere to a distant observer with the \gy code.   Here, we present and test our fully relativistic numerical pipeline. To discuss and illustrate the importance of realistic atmosphere models we compare our model spectra to simpler models like the commonly used isotropic color-corrected blackbody emission.
We highlight the importance of considering realistic model-atmosphere spectra together with relativistic ray tracing to obtain accurate predictions.
We also insist on the crucial impact of the star's rotation on the observables.
{Finally, we close a controversy that has been appearing in the literature in the recent years regarding the validity of the \at code.}
\end{abstract}


\section{Introduction}
Neutron stars create extreme environments that harbor forms of ultra-dense matter that cannot be reproduced on Earth.   The very complex properties of the stellar matter are encapsulated in the equation of state (EoS) that allows to close the system of equations describing the star's equilibrium.   This EoS is the link between the nuclear physics in the interior of the neutron star and astrophysical observables.   In particular, a specific EoS will only allow certain pairs of values of neutron star mass $M$ and radius $R$~\citep[see e.g.][]{haensel07}.   Constraining the pair $(M,R)$ for a particular neutron star thus gives access to the nature of the EoS~\citep[for a recent review see][]{ozel16b}.  {We note that the star's rotation has an impact on the $M$-$R$ relation and that
the relation that is generally represented in typical figures is obtained for non-rotating stars.}

Neutron star masses and radii may be constrained by various methods, such as
\begin{itemize}
\item the modeling of pulse profiles for sources that show an oscillating light curve most probably
      due to a rotating hot spot on the neutron star's surface;
\item the analysis of kHz quasi-periodic oscillations seen in the neutron star's power density spectrum;
\item the spectroscopic study of neutron stars in low-mass X-ray binaries (LMXB) that show so-called type-I
      X-ray bursts, i.e. a sudden burning of the fuel accreted from the secondary star;
\item the modeling of thermal spectra from cooling isolated neutron stars or thermal dominated spectra of neutron stars in quiescent LMXB (qLMXB).
\end{itemize}
For reviews of these methods and of their limitations see e.g.~\citet{miller13, miller16, ozel16b,haensel16}.
In this article we focus on the spectral methods, and more specifically on type-I bursters. 
This choice is in particular motivated by the fact that
LMXB tend to have magnetic fields of the order of $10^{8 - 9}$~G, which is weak for a neutron star.
It is thus fair to assume that the magnetic field does not influence the radiative properties of the atmosphere~\citep{miller13},
which greatly simplifies the problem.
We note that the numerical methods we present here are as well adapted for qLMXB, and
could be easily developed in future works to model pulse profiles (in a scarcely magnetized environment).

Type-I bursters, in particular when they show photospheric radius expansion \citep[PRE, see e.g.][]{fujimoto86, vanparadijs87},
have been used to constrain masses and radii of neutron stars ever since the method was proposed by~\citet{vanparadijs79}.
The interpretation of observational data in the framework
of this method may suffer from a series of limitations that can be divided into two main issues:
\begin{itemize}
\item the correct modeling of the emitted spectra;
\item the way the effects of strong gravity are taken into account.
\end{itemize}
We will now review the recent progress in these two directions.

Regarding the modeling of the emitted spectra, it is first possible
that not all the emission comes from
the surface of the star (but for instance from residual accretion),
and that not the whole surface of the star is emitting radiation~\citep{miller16}.
We will assume in the remaining of the article that the full surface of the star,
and it only, emits radiation. The main source of difficulty is then to model
the emission arising from the star's atmosphere.

The best constraints obtained from spectral methods use data recorded by the
Rossi X-ray Timing Explorer (RXTE), using color-corrected blackbody emission
at the surface of the star~\citep{ozel09,ozel16}.
Having a huge effective area, RXTE was able to collect a sufficient
number of counts within milliseconds, allowing to make proper spectral analysis.
Nevertheless, using a color-correction factor is always a simplified treatment
which boils down to only shifting the blackbody
radiation to mimic the emission from a true atmosphere.
In the recent years, realistic atmosphere models have been developed
to study neutron stars in type-I bursters and estimate their masses and radii.

There exist two classes of neutron star atmosphere models in the literature.
The first class are model atmospheres which are approximate for the reason that they oversimplify the process of electron scattering~\citep{heinke06, webb07, guillot11}.
This class of neutron star model atmospheres is available in the widely used {\sc xspec}~\citep{arnaud96} software fitting package.
The second class of available model atmospheres does include accurate Compton scattering process,
which is critically important to model atmospheres of hot neutron stars (with effective
temperature of a few times $10^7$~K, characteristic of X-ray bursters) and their
spectra~\citep{madej89,madej1991a,madej1991b,madej04,majczyna05,suleimanov11,suleimanov12,suleimanov16}.
Those models were successfully used for the mass and radius determination
of non-rotating NS~\citep[][and references therein]{majczyna05AcA,kusmierek11,suleimanov16}.
We would like to point out here that only model atmospheres with Compton scattering fully taken into
account can be used for mass and radius determination. Recently we
proved that the Compton redistribution functions that we use in the radiative transfer equations
are computed with excellent precision~\citep{madej17}.

Regarding the impact of strong
gravity, studies have been devoted to its effect on the radiative transfer in
the atmosphere itself, as well as on the subsequent propagation of
photons towards the distant observer.
Most spectroscopic studies that aim at constraining neutron stars' properties from thermal X-ray spectra
of type-I bursters approximate the effects of general relativity
by a single scalar factor, the surface gravitational redshift $z$. Such a scalar correction
is sufficient to represent relativistic effects provided the thickness of the neutron star's atmosphere
is small compared to the radius of the star. Only recently, \citet{medin16} presented
models of geometrically thick neutron-star atmospheres with accurate Compton scattering, taking into account the
relativistic bending in the radiative transfer process itself, within the neutron star's atmosphere.

The effects of strong gravity on the subsequent propagation of photons from the atmosphere
to the distant observer has been the subject of a series of studies in the past decade.
\citet{cadeau05,cadeau07} compared ray-tracing of photons in an approximate analytical spacetime with no shape modification of the star to ray-tracing of photons in an exact numerical neutron star spacetime~\citep{stergioulas95} with the deformation of the surface taken into account.   These studies lead to the conclusion that the oblate shape of the rotating star is the most important factor that should be considered.   A recent series of papers by~\citet{baubock12, baubock13, baubock15} studied the importance of the rotation of the star as well as the quadrupole moment of the metric considering an analytical approximation to the exact neutron star metric, namely the~\citet{hartle68} metric.   The validity of this analytical approximation has been studied by~\citet{berti05, baubock13} and showed a satisfactory agreement for most astrophysical contexts.
%
%

{An interesting very recent series of papers by~\cite{nattila17a,nattila17b} is investigating
in the same direction as our paper. It describes a new open-source ray-tracing code for
neutron-star observables computation in approximate spacetimes, and presents the first
direct fit to bursting neutron star data of model atmospheres (as contrasted to all previous fits
performed with modified blackbodies). It is obviously of crucial importance that more than
one numerical pipeline like ours should be published so that cross-checks of the rather involved
outputs can be made.}

{Recent confirmation of the origin of short gamma-ray bursts as mergers of
binary neutron-star systems \citep{GW170817,GW170817MMA,GW170817GRB} creates a
new method of establishing neutron stars' parameters. From the observations of tidal
deformation imprinted on the waveform's phase during the last orbits of the inspiral,
one can distill the information on the size of the star
and the stiffness of the EoS. The analysis performed in \citet{GW170817} suggests
that the component stars taking part in the merger had radii most likely smaller
than 14 km at the measured mass range around $1.4\ M_\odot$.}

The aim of this article is to present a numerical pipeline that implements the most accurate simulated neutron star spectrum to date, taking into account at the same time a precise atmospheric model, and considering all general-relativistic effects of the star's geometry on the photon propagation.   We compute an accurate numerical metric that in particular takes into account the exact deformation of the star due to its rotation.   We solve the hydrostatic and radiative transfer problem in the neutron star's atmosphere incorporating the exact value of the local (varying) surface gravity at the star's surface.   This local spectrum is then transported to a distant observer by means of a ray-tracing code that adheres to the exact numerical metric of the neutron star.   An obvious interest of such a numerical pipeline is to provide a practical testbed for approximate methods, that are faster and more adapted for fitting data.   However, our model can also be used to produce tables of model spectra that could be implemented in
{\sc xspec} and used for fitting data.   Indeed, we will see below that the computing time necessary to produce a spectrum is reasonable for this prospect ($\approx 5$~min on a $2.9$ GHz Intel Core i7 processor of a personal laptop to produce one spectrum).   Moreover, such a pipeline also offers the possibility to investigate in full details the impact of the EoS on the resulting spectrum, as well as to accurately predict the observables associated to "extreme" systems that have not yet been observed to date: very fast-rotating
and very massive neutron stars. We highlight that our pipeline is for the time being only partially open-source, but we aim at
making it fully open-source in the close future.

The numerical method is presented in Section~\ref{sec:methods}, and the results of the simulations in Section~\ref{sec:simus}.
Section~\ref{sec:conc} summarizes our work and gives future directions.
{Appendix~\ref{appendix} demonstrates the validity of the \at code, that has been
questioned in recent articles.}

\section{Description of the methods}
\label{sec:methods}
\begin{figure}
   \centering
   \includegraphics[width=\columnwidth]{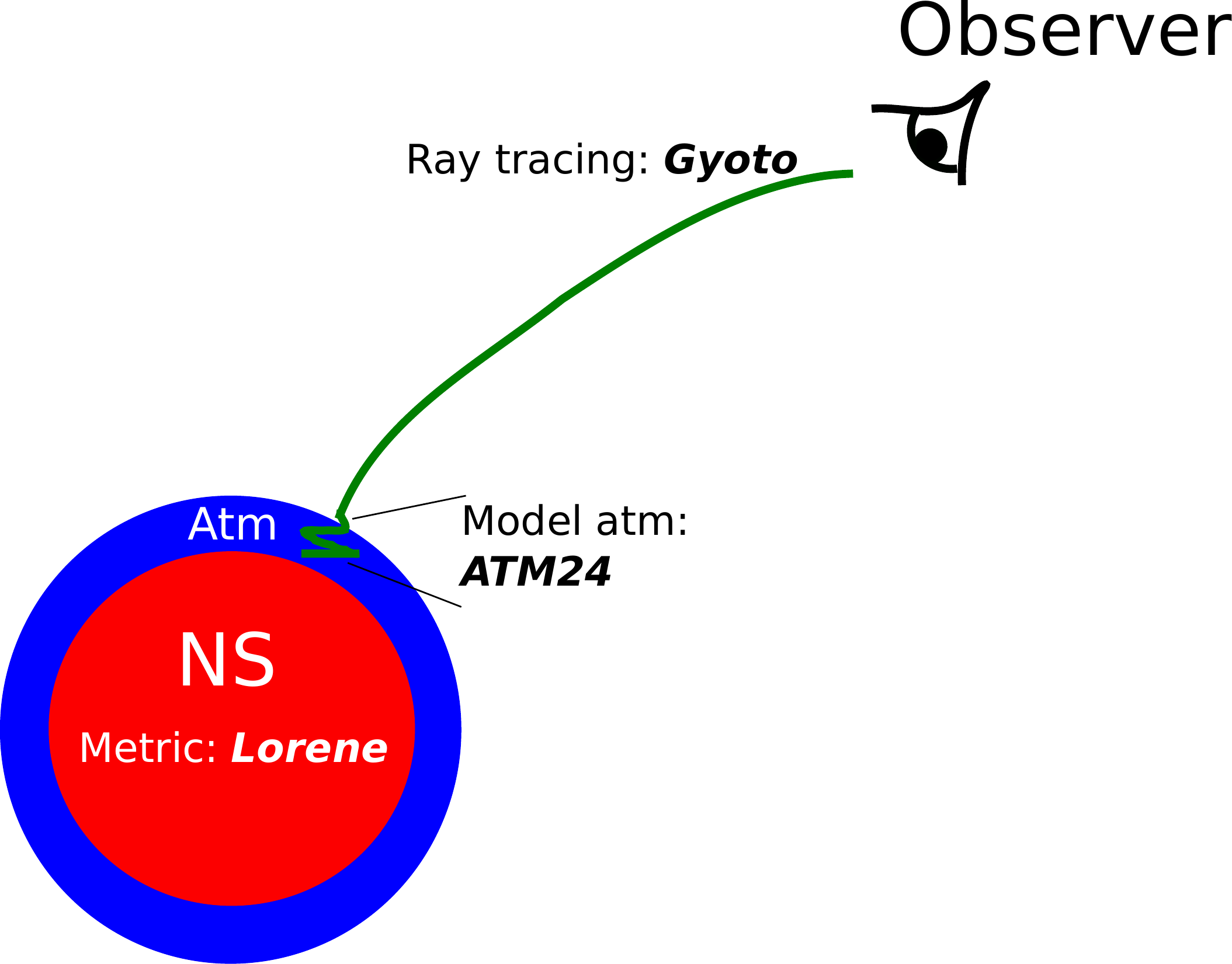}
      \caption{Sketch of the methods used to compute images and spectra of neutron stars.   The neutron star is depicted in red, its structure and spacetime metric is computed with the \lonr code.   Its atmosphere is represented in blue (not to scale, the atmosphere is extremely thin as compared to the star's radius, typically $10$s of cm compared to $10$~km).   The radiative transfer equation is solved there together with hydrostatic equilibrium with the \at code.   Finally, the emitted photons are ray-traced to a distant observer using the \gy code which incorporates the neutron star's metric computed by \lonr.}
      \label{fig:sketch}
   \end{figure}

Images and spectra of neutron stars are computed following the sketch of Figure~\ref{fig:sketch}.
This Section describes the various pieces that appear in that sketch.

\subsection{Spacetime metric}
\label{sec:spacetimemetric}
To describe the spacetime generated by the neutron star we adopt the quasi-isotropic coordinates $(t,r,\theta,\pp)$. For simplicity we assume stationarity and axisymmetry, with Killing vectors $ \boldsymbol{\partial}_t$ and $ \boldsymbol{\partial}_\pp$ corresponding to time and rotational symmetries.
Under the assumption of pure rotational fluid motion about the symmetry axis,
the general metric element reads
\bea
\dd s^2 = g_{\mu\nu}\, \dd x^\mu \dd x^\nu &=& -N^2 \dd t^2 + A^2 \left( \dd r^2 + r^2 \dd \theta^2 \right) \nn \\
&&+ B^2 r^2 \sin^2 \theta \left( \dd \pp - \omega \dd t \right)^2, \label{eq:metric}
\eea
where $N$, $A$, $B$ and $\omega$ are four functions of $(r,\theta)$,
$N$ being called the \emph{lapse function}.
The fluid 4-velocity $\mathbf{u}$ is a linear combination of the two Killing vectors (pure
rotational motion hypothesis):  $\mathbf{u} = u^t \left( \boldsymbol{\partial}_t + \Omega \,\boldsymbol{\partial}_\pp \right)$, where the contravariant time component of the 4-velocity
$u^t$ is expressible as $u^t = \Gamma/N$, with $\Gamma$ the Lorentz factor of the fluid with respect to the zero angular momentum observers (ZAMO). In the following we will assume rigid rotation, i.e. a constant value of the star's angular velocity $\Omega = u^\phi/u^t = \dd\phi/\dd t$.

The metric of a neutron star rigidly rotating with angular frequency $\Omega$ is fully fixed once the dense matter equation of state
inside the star is selected and the central density $\rho_c$ (or pressure $P_c$) is chosen. This is a natural choice
for computational purposes which allows to integrate the general relativistic equations of structure for a uniformly rotating body.
Two out of the three following global quantities, which are well defined in general relativity
for a rotating star, also uniquely determine a rotating configuration and the properties of the
metric at the star's surface:
\begin{itemize}
  \item the gravitational (ADM, \citealt{ADM}) mass $M$,
  \item the baryon mass $M_b$,
  \item the total angular momentum $J$.
\end{itemize}
For the relations between these global parameters, properly defining sequences of stellar configurations,
see e.g.~\cite{Zdunik2006}. Another important quantity is the circumferential stellar radius $R$, defined such that the circumference of the star in the equatorial plane, as
given by the metric tensor, is $2\pi\,R$. Then $R$ is
related to the equatorial
quasi-isotropic coordinate radius $r_\mathrm{eq}$ by the metric function $B(r,\theta)$,
according to $R = B(r_\mathrm{eq},\pi/2) \,r_\mathrm{eq}$.

In order to obtain the accurate solutions for rotating neutron stars configurations at arbitrarily high spin (i.e., beyond the slow-rotation approximation), we employ the \textsc{nrotstar} code \citep{gourgoulhon10} built using the free and open-source \lo library~\citep[][\url{http://www.lorene.obspm.fr}]{lorenelib},
which provides spectral-methods solvers to the Einstein equations. The \textsc{nrotstar} code allows for fast computation of the stellar surface, described by the coordinate radius $r_\mathrm{surf}(\theta)$ as a function of $\theta$, the 4-velocity of the fluid at the surface, as well as the fluid 4-acceleration at the surface.
In all the star, the fluid 4-acceleration $\mathbf{a}$ has covariant components
given by
\be
\label{eq:accel}
a_\mu = u^\nu \nabla_\nu u_\mu = -\partial_\mu \left( \mathrm{ln}\, u^t\right).
\ee
The second expression results from the assumptions of circularity and rigid rotation, i.e.
from the fact that the fluid 4-velocity is
$\mathbf{u} = u^t \mathbf{k}$, where
$\mathbf{k} = \boldsymbol{\partial}_t + \Omega \boldsymbol{\partial}_\pp$
is a Killing vector, since $\Omega$ is constant.
The surface gravity $g_{\rm s}$ is the norm of $\mathbf{a}$ at the stellar surface:
\be
g_{\rm s} = \sqrt{a_\mu a^\mu} = A \sqrt{\left(a^r\right)^2 + r^2 \left(a^\theta\right)^2},
\ee
where the second equality results from the metric (\ref{eq:metric})
and the fact that Eq.~(\ref{eq:accel}) along with the hypothesis of stationarity
and axisymmetry imply $a_t = a_\pp = 0$.

The first integral of the relativistic Euler equation governing the fluid motion
is $h/u^t = \mathrm{const}$~\citep[see e.g.][]{gourgoulhon10}, where $h$ is
the fluid's relativistic specific enthalpy.
At the surface of the star, the fluid's internal energy
and pressure tend to zero, so that $h=1$. It follows from the first integral
that the stellar surface is an isosurface of $u^t$. From Eq.~(\ref{eq:accel}), we conclude
that the fluid 4-acceleration gives the normal to the stellar surface
(which is a timelike hypersurface in spacetime).
 For some photon escaping the star with a tangent null 4-vector $\mathbf{p}$, the emission angle $\epsilon$ between the direction of photon emission in the frame corotating with the star and the local normal reads
\be
\label{eq:angle}
\cos \epsilon = - \frac{\mathbf{n} \cdot \mathbf{p}}{\mathbf{u} \cdot \mathbf{p}}
\ee
where $\mathbf{n} = \mathbf{a}/g_{\rm s}$ is the unit spacetime normal to the stellar
surface, $\mathbf{u}$ is the surface value of the fluid's 4-velocity
and a dot denotes the scalar product taken with the metric (\ref{eq:metric}).
As we will see below, all these quantities are necessary for the ray-tracing and the atmosphere radiative transfer calculations.

\subsection{Directional atmospheres}

{ We compute local spectra emitted in the star's hydrogen/helium atmosphere via the radiative transfer code {\at}
(Madej et al. 2018, in preparation). Previous versions of the code were described in~\citet{madej04,majczyna05}.  
We note that the neutron star's atmosphere has a very
small height ($\approx 1$~m as compared} to the star's radius of $\approx 10$~km) so that the radiative transfer computation is essentially local.
This allows us to neglect relativistic effects in this Section and use a Newtonian treatment.

The equation of radiative transfer in a plane-parallel atmosphere
can be written in the following form:
\begin{eqnarray}
 \label{equ:tra}
 & \mu & {\partial I_\nu (z,\mu) \over {\rho \partial z}} =
 \kappa^\prime_\nu (1-e^{-h \nu/kT})(B_\nu-I_\nu)  \\ & + &
 \left( 1+{c^2 \over {2h \nu^3}} I_\nu \right)
 \oint \limits_{\omega^\prime } {d\omega^\prime \over 4\pi}
\int \limits _{0}^{\infty} {\nu \over {\nu ^\prime }} \sigma (\nu ^\prime
\rightarrow \nu , \vec n {^\prime} \cdot \vec n )  I_{\nu ^\prime}
(z, \vec n^\prime ) d\nu ^\prime   \nonumber  \\  &- &\,
I_\nu (z, \mu ) \oint \limits_{\omega ^\prime} {d\omega^\prime \over {4\pi}}
\int \limits _{0}^{\infty}  \sigma (\nu \rightarrow \nu ^\prime ,
\vec n \cdot \vec n {^\prime} )  (1+ {c^2 \over {2h{\nu ^\prime }^3 }}
I_{\nu ^\prime} ) \, d\nu ^\prime  \, . \nonumber
\end{eqnarray}
assuming that sources of true  absorption ($\kappa_\nu$) and thermal
emission ($j_\nu = \kappa_\nu B_\nu$) are in local thermodynamic equilibrium (LTE).
Here, $z$ is the geometrical depth, and $\mu = \cos \epsilon$.
Eq.~(\ref{equ:tra}) does not include relativistic corrections
on the neutron star surface, which will be treated in the next Section.
The above equation of transfer was a starting point
in earlier investigations \citep{sampson59,madej89,madej1991a,madej1991b}.

Variable $\sigma_\nu$ denotes the coefficient of Compton scattering
at a given frequency $\nu$ integrated over all incoming frequencies
$\nu^\prime$. The variable $\sigma (\nu \rightarrow \nu ^\prime ,
\vec n \cdot \vec n {^\prime} $) denotes the differential Compton scattering
cross section, and $\kappa_{\nu}^{\prime}$ is the absorption coefficient
uncorrected for stimulated emission. All the opacity coefficients
are given for 1 gram of matter (units of cm$^{2}$ g$^{-1}$).
The relation between the differential Compton scattering cross section and
the integrated Compton scattering coefficient is given by the following
equation:
\begin{equation}
\sigma_\nu=\oint \limits_{\omega ^\prime} {d\omega^\prime \over {4\pi}}
\int \limits _{0}^{\infty} \sigma (\nu \rightarrow \nu ^\prime ,
\vec n \cdot \vec n {^\prime})
 d\nu^\prime.
\end{equation}

Compton scattering differential cross section must
fulfil the relation:
\begin{equation}
\sigma ( \nu \rightarrow \nu {^\prime }, \vec n \cdot \vec n {^\prime} )
\, \nu ^2  e^{-h\nu /kT} = \,
\sigma ( \nu {^\prime }\rightarrow \nu , \vec n {^\prime} \cdot \vec n )
\, {\nu ^\prime }^2 e^{-h\nu ^\prime /kT}  \> ,
\label{equ:sig}
\end{equation}
which results from the detailed balance of this process in global
thermodynamic equilibrium~\citep{pomraning73}.

{
Moreover, we define new variables:
\begin{equation}
\kappa _\nu = \kappa _\nu {^\prime } \, (1-e^{-h\nu /kT})
\end{equation}
and
\begin{equation}
\sigma (\nu \rightarrow \nu ^\prime,\vec n \cdot \vec n ^\prime ) \,
= \sigma_\nu \, \phi (\nu , \nu ^\prime,\vec n \cdot \vec n ^\prime )
\> ,
\end{equation}
where the Compton scattering redistribution function $\phi$ is normalized 
to unity:
\begin{equation}
\oint \limits_{\omega ^\prime } {{d{\omega ^\prime }} \over 4\pi }
\int \limits_{0}^{\infty } \phi (\nu ,\nu ^\prime,\vec n \cdot \vec n ^\prime )
\, d{\nu ^\prime } =1 \, .
\end{equation}
Angle-dependent Compton scattering redistribution function 
$\phi$ was approximated by its zeroth angular moment, 
\begin{equation}
\Phi (\nu , \nu ^\prime )=\oint \limits_{\omega ^\prime }
\phi (\nu ,\nu ^\prime ,\vec n \cdot \vec n {^\prime} ) \,
{d{\omega ^\prime }\over 4\pi } \, .
\end{equation}
Then, we write the equation of transfer (\ref{equ:tra}) on the 
monochromatic optical depth scale 
$d \tau_\nu=-(\kappa_\nu + \sigma_\nu)
\rho dz$:
\begin{eqnarray}
  \label{equ:tau}
 \mu \, {\partial I_\nu \over {\partial \tau_\nu }} & = &
I_\nu - \epsilon _\nu \, B_\nu	\> -(1-\epsilon_\nu)J_\nu \\
& + & (1-\epsilon_\nu ) J_\nu \,
\int\limits_{0}^{\infty}  \Phi (\nu , \nu ^\prime )
\, (1+ {c^2 \over {2h{\nu ^\prime }^3 }} J_{\nu ^\prime} ) \,
d\nu ^\prime \>  \nonumber \\
& - &(1- \epsilon_\nu ) \, ( 1+ {c^2 \over {2h\nu ^3}} J_\nu ) \,
\int\limits_{0}^{\infty} \Phi (\nu ,\nu ^\prime) J_{\nu ^\prime} \nonumber
\\ &\times & \, {\left( {\nu \over {\nu ^\prime }} \right) } ^3
\exp \left[ -{{h(\nu - {\nu ^\prime }) }\over {kT}} \right]
\, d\nu ^\prime  .  \nonumber
\end{eqnarray}
Dimensionless monochromatic absorption is defined as 
$\epsilon_\nu = \kappa_\nu / (\kappa_\nu + \sigma_\nu)$.
We point out here that the equation of transfer, Eq.~(\ref{equ:tau}), is
not linear and is quadratic with respect to the unknown $J_\nu$.

{ {Here we stress that the equation of transfer assumes that Compton 
scattering is an isotropic process. 
The angle-integrated normalized redistribution function $ \Phi (\nu ,\nu ^\prime)$ was derived 
from the exact redistribution function $R_1$ as presented in \citet{madej17}. }

The above equation of transfer is accompanied by two boundary conditions,
the first being:
\begin{equation}
{d \over {d \tau_\nu }} \, (f_\nu J_\nu )\, = H_\nu (0)
\label{equ:edd} 
\end{equation}
at the top of the model atmosphere ($\tau_\nu=0$). At the bottom
we require $J_\nu=B_\nu$, according to the usual thermalization condition.  
The variable $f_\nu$ denotes the well-known Eddington factor. 

In this paper we solve the above equation precisely iterating 
Eddington factors $f_\nu$ at all frequencies and optical depth points. 
In such a way, we exactly reproduce the angular behavior of the radiation
field, and determine the angular stratification of the specific intensity 
$I_\nu (\tau,\mu)$. 
}

Our equations  and the Compton redistribution functions
\citep{nagirner93,suleimanov12}
work correctly for both large and small energy
exchange between X-ray photons and free electrons at the time of
scattering.
{The resulting} model atmosphere and theoretical spectra represent the most accurate
quantum mechanical solution of the Compton scattering problem, as shown
in our recent paper \citet{madej17}.  The above equations produce accurate
solutions also when the initial photon energy before
or after scattering exceeds the electron rest mass ($m_e c^2 = 511 $ keV).

In our numerical procedure, the equation above is solved with the structure of the gas kept in
radiative, hydrostatic, and ionization equilibrium
simultaneously. Condition of radiative equilibrium implies on each depth
level $\tau_\nu$ that
\begin{equation}
 \int\limits_0^\infty H_\nu (\tau_\nu) \, d\nu
  = {\sigma_{R} T_{\rm eff}^4 \over {4\pi}}  \, ,   \label{equ:lev}
\end{equation}
where $\sigma_{R} = 5.66961 \times 10^{-5}$ (in cgs units). The second condition of hydrostatic equilibrium can be expressed in the
usual form
\begin{eqnarray}
  {dP_g \over {d\tau}}  &=&  {g_{\rm s}\over{(\kappa+\sigma)_{std}}}
    - {dP_r \over {d\tau}}  \\ &=& {g_{\rm s}\over{(\kappa+\sigma)_{std}}}
- {4\pi \over c} \, \int \limits _{0}^{\infty}
  { \kappa_\nu+\sigma_\nu \over {(\kappa_\nu+\sigma_\nu)_{std}}}\, H_\nu  \, d\nu 
\, \nonumber
\end{eqnarray}
where $g_{\rm s}$ is the surface gravity, which is computed self-consistently along the neutron star's surface with \lonr.   The standard opacities and standard optical depth $\tau$ correspond to the same
variables at the arbitrarily fixed frequency.
The hydrostatic equation is coupled to the radiative transfer by the mass density $\rho$
that appears in the optical depth $\tau_\nu$.
{We use the equation of state of an ideal gas and compute ionization and
excitation populations, therefore absorption coefficient $\kappa_\nu$,
using Saha and Boltzmann equations (assuming local thermodynamic equilibrium). }

The inputs of \at thus are:
\begin{itemize}
  \item the composition of the atmosphere, which is assumed here to be composed of hydrogen and helium, in solar abundance,
  \item the surface effective temperature, assumed here to be $T_\mathrm{eff} = 10^7$~K,
  \item the surface gravity $g_{\rm s}$, self-consistently computed by \lonr along the surface.
\end{itemize}
The output is the local spectrum of emitted specific intensity $I_\nu(\nu,\mu)$, as a function
of emitted frequency and cosine of emission angle.
This output can also slightly depend on the polar angle $\theta$ connected to the star's shape, given that the surface gravity becomes a function of $\theta$ for rotating stars, which we fully take into account in this paper computing
the accurate distribution of $I_\nu(\nu,\mu,\theta)$.

{We note that the validity of the \at code has been recently put in question
by \cite{suleimanov12}, as well as more recently by~\cite{medin16}, on the basis
that one model spectrum obtained by these authors differs from that computed by \at.
We show in Appendix~\ref{appendix} that the
 \at spectrum used for comparison simply was not properly converged, as was suggested
in the discussion of \cite{suleimanov12}. Appendix~\ref{appendix} shows that fully
converged \at spectra essentially do agree with \cite{suleimanov12}.}

\subsection{Ray tracing}
The open-source ray tracing code \gy~\citep[see][and \url{http://gyoto.obspm.fr}]{vincent11,vincent12} is used to compute null geodesics backwards in time, from a distant observer towards the neutron star.   Photons are traced in the neutron star's metric as computed by the \lonr code.   Once the neutron star's surface is reached, the emission angle $\epsilon$ between the photon's direction of emission and the local normal is evaluated following Eq.~(\ref{eq:angle}).   Given the observed photon frequency $\nu_\mathrm{obs}$, the emitted frequency $\nu_\mathrm{em}$ is found knowing the redshift which is defined by the local value of the 4-velocity at the neutron star's surface and which is computed by \lonr.   Finally, the local surface gravity is also known from the \lo output. Consequently, the local value of the emitted specific intensity $I_\nu^\mathrm{em}(\nu_\mathrm{em}, \mu, g_{\rm s}(\theta_\mathrm{em}))$ at a polar angle $\theta_\mathrm{em}$, can be interpolated from the output table computed by the \at code.   The observed specific intensity is then $I_\nu^\mathrm{obs} = \nu_\mathrm{obs}^3 / \nu_\mathrm{em}^3\,I_\nu^\mathrm{em}$.   The map of observed specific intensity (i.e. the image) in the frame of a distant observer is thus at hand.   Performing such computations for a set of observed photon frequencies and summing these images over the observer's screen pixels (i.e. over the directions of photon reception) allows to obtain an observed spectrum $F_\nu^\mathrm{obs}$.

\section{Stellar and atmospheric models, ray-traced images and spectra}
\label{sec:simus}

\subsection{Stellar models}


We consider two models of neutron stars described by the SLy4 dense-matter equation of
state \citep{chabanat98,douchin01}, supplemented with the description of the neutron star's
crust of \citet{HaenselP1994}: a non-rotating configuration and a configuration
rotating rigidly at 716 Hz (matching the frequency of currently
fastest-spinning pulsar PSR J1748-2446ad, \citealt{Hessels2006}). For both
configurations we select a gravitational mass of $M = 1.4\ M_\odot$. In
general, the surface gravity of a non-rotating neutron star is
\be
  g_{{\rm s},0\, {\rm Hz}} = \frac{GM}{R^2}\frac{1}{\sqrt{1 - 2GM/Rc^2}}.
\label{eq:gsnrot}
\ee
For a mass of $M=1.4\ M_\odot$ and considering the SLy4 EoS,
$g_{{\rm s},0\, {\rm Hz}} =  1.68\times
10^{14}\,\mathrm{cm}\,\mathrm{s}^{-2}$ (see \citealt{BejgerH2004} for a summary
for other EoS models). In contrast, a rotating neutron star has a surface gravity varying along the
$\theta$ direction, as described in Sect.~\ref{sec:spacetimemetric}.
In Fig.~\ref{fig:surfgrav} we compare the surface gravities for both our models.
The resulting values of surface gravities and circumferential radii are
of order $g_{\rm s} = 1.5\times 10^{14}\,\mathrm{cm}\,\mathrm{s}^{-2}$ and $R = 12$~km.
In comparison, the neutron star model often considered as a reference,
with $M = 1.4\ M_\odot$ and R = 10 km, yields
$g_{\rm s} = 2.43 \times 10^{14}\,\mathrm{cm}\,\mathrm{s}^{-2}$ \citep{BejgerH2004}.

   \begin{figure}
            \includegraphics[width=\columnwidth]{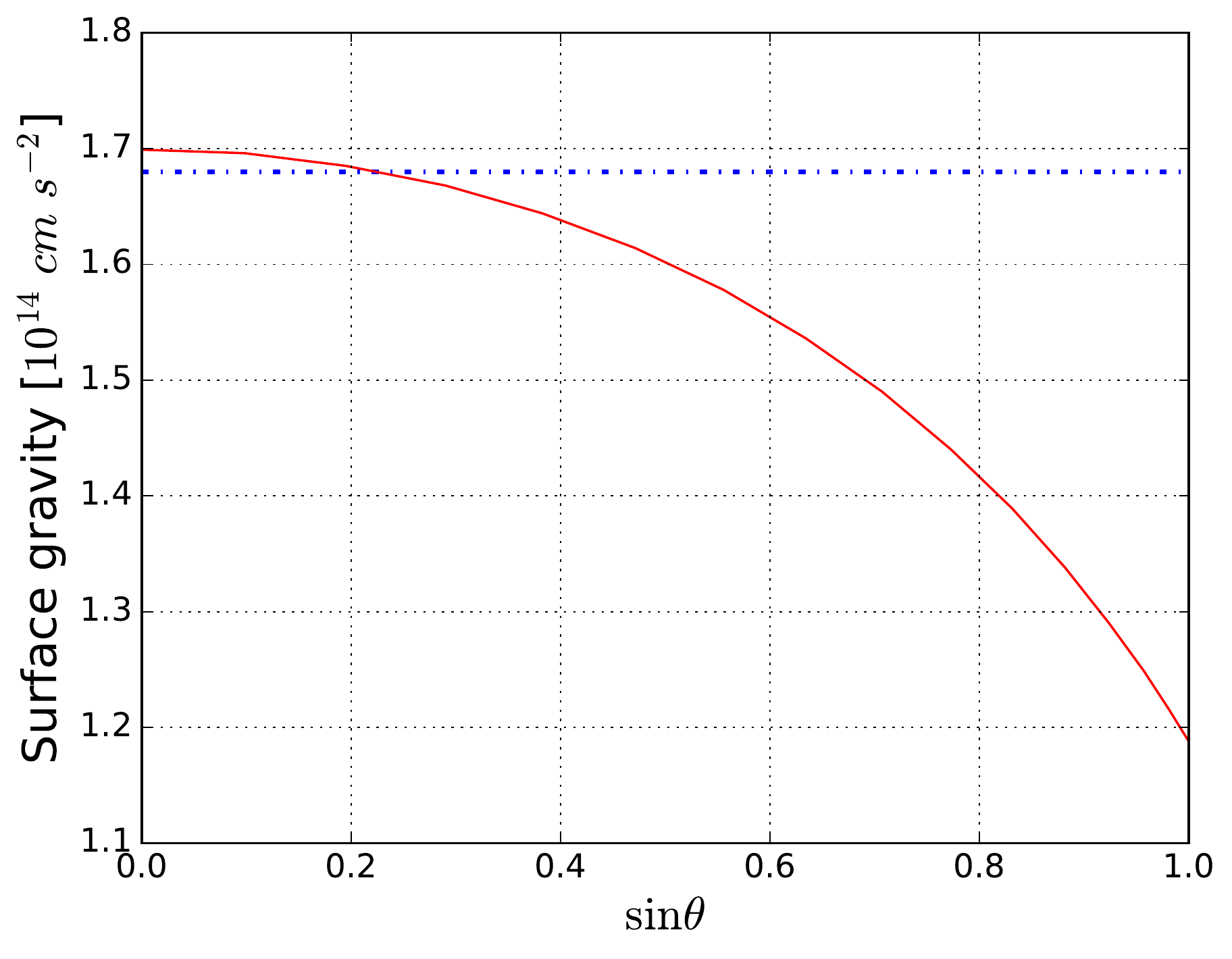}
      \caption{Comparison of surface gravities between a non-rotating neutron star (blue
dash-dotted line at a constant value of $1.68\times
10^{14}\,\mathrm{cm}\,\mathrm{s}^{-2}$) and for a neutron star rigidly rotating at 716 Hz
(red line, with values between $1.7\times 10^{14}\,\mathrm{cm}\,\mathrm{s}^{-2}$
for the pole at $\sin\theta=0$ and $1.19\times 10^{14}\,\mathrm{cm}\,\mathrm{s}^{-2}$ for the
equator at $\sin\theta=1$). In both cases the gravitational mass equals $1.4\ M_\odot$.}
\label{fig:surfgrav}
   \end{figure}

We also compare the surface figures of stars. Figure~\ref{fig:surfaces}
shows them in coordinate values for both selected configurations and
compare them with the slow-rotation approximation to the shape of the surface:
\be
  r\left(\theta\right) = r_{\mathrm{eq}} - \left(r_{\mathrm{eq}} - r_p\right)\cos^2\theta,
\label{eq:resr}
\ee
where $r_{\mathrm{eq}}$ and $r_p$ are the equatorial and polar quasi-isotropic radii.
{This equation is used for instance in the Hartle-Thorne metric
used by~\cite{baubock12}.}
Assuming that the
polar and equatorial radii are known, the error introduced by using the
slow-rotation approximation at the rotational frequency of 716~Hz and mass
$1.4\ M_\odot$ for the SLy4 EoS is maximally about 100~m in radius (for
comparison, the maximal difference at 900~Hz is about 0.5~km).

   \begin{figure}
            {\hspace*{0.018\columnwidth}
            \includegraphics[width=0.97\columnwidth]{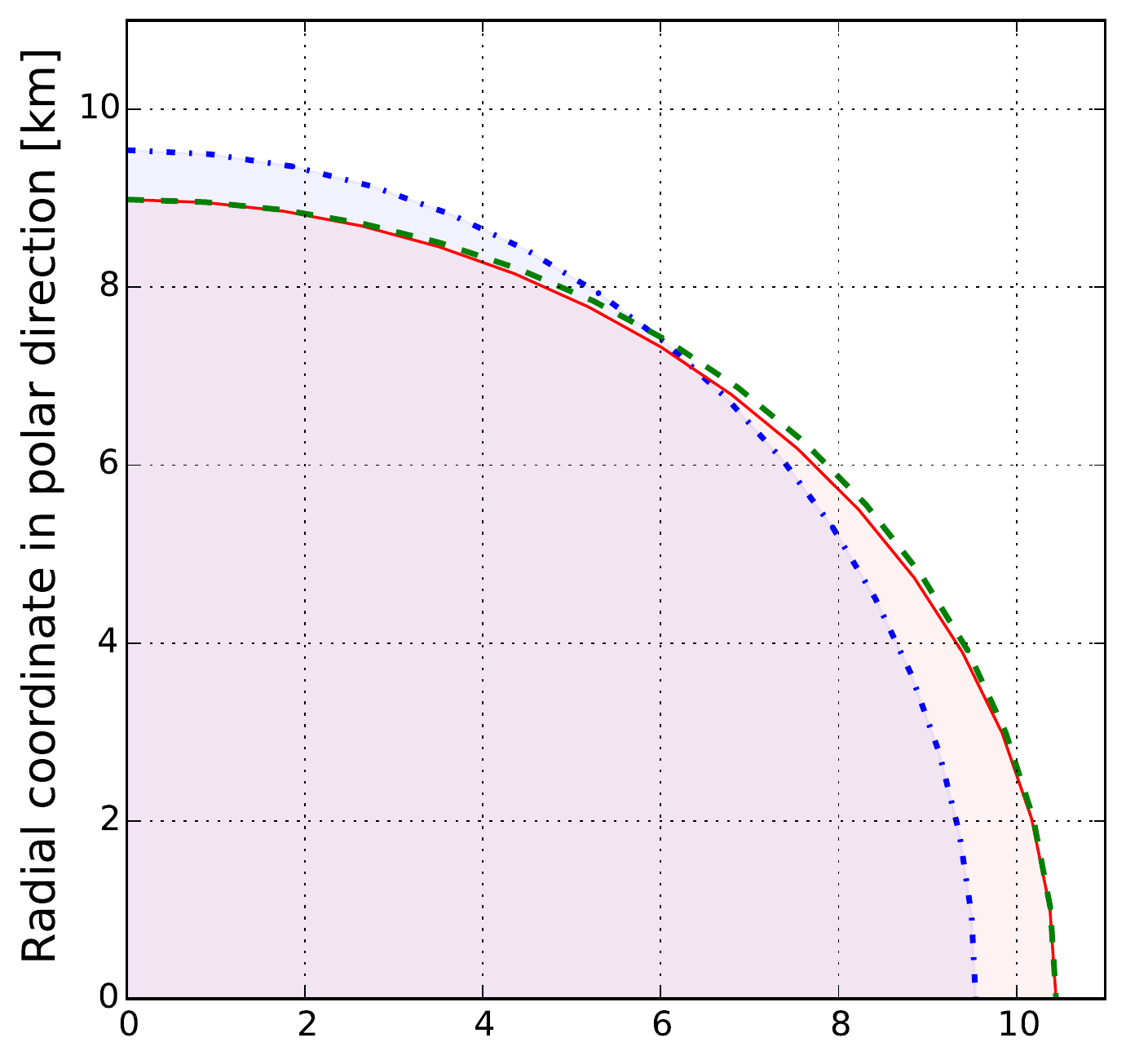}}
            \vskip 0.1cm
            {\includegraphics[width=\columnwidth]{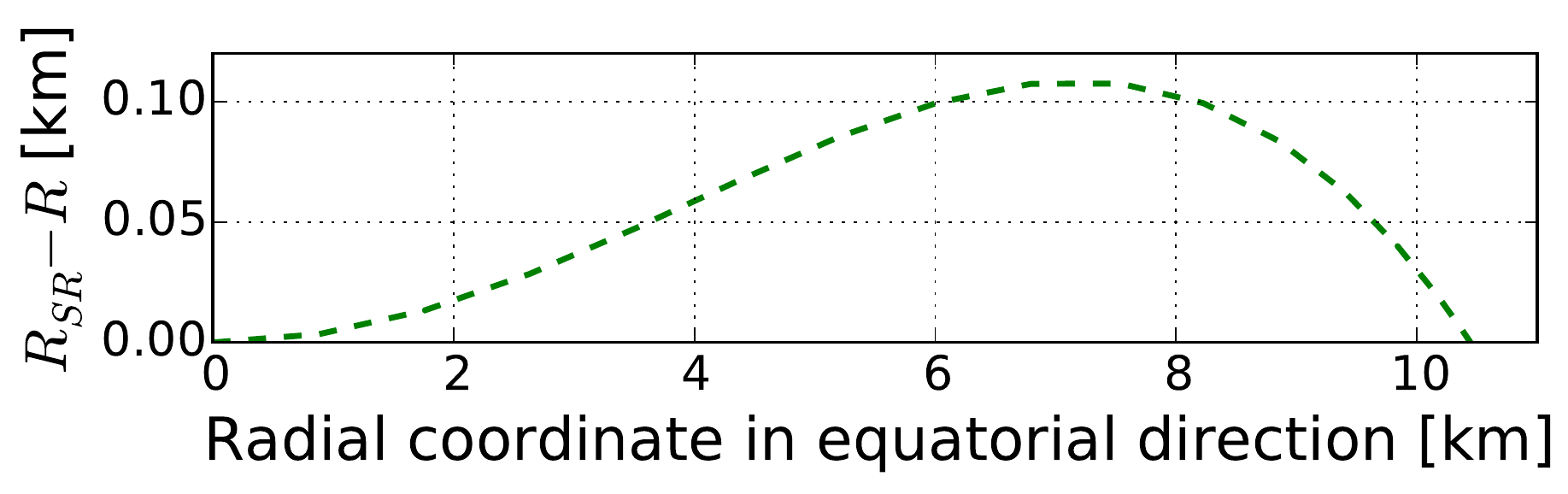}}
      \caption{Surface figures for selected configurations in coordinate values.
      The blue dash-dotted line denotes the non-rotating star (circumferential
      radius $R_{0\,\mathrm{Hz}} = 11.72$~km). The red line denotes the surface of the star rotating at 716~Hz
      (equatorial circumferential radius $R_{716\,\mathrm{Hz}} = 12.65$~km, $r_p/r_{\mathrm{eq}} = 0.86$).
      The green dashed line is the slow-rotation approximation of Eq.~(\ref{eq:resr}).
      The lower panel presents the difference between the shape of the rotating star obtained
      with \lonr, and the slow-rotation approximation of Eq.~(\ref{eq:resr}).}
      \label{fig:surfaces}
   \end{figure}

The spacetime generated by a rotating star in full general relativity is different from the slow-rotation
approximation. Figure~\ref{fig:metricfunctions} shows the comparison of the metric function lapse $N$ with
its approximation $1 - M/r$, and the frame-dragging metric term (shift vector component $\omega = -\beta^\phi$)
with $2J/r^3$, where $J$ is the total angular momentum of the configuration rotating at 716 Hz
(see also \citealt{gourgoulhon10}).

   \begin{figure}
      \includegraphics[width=\columnwidth]{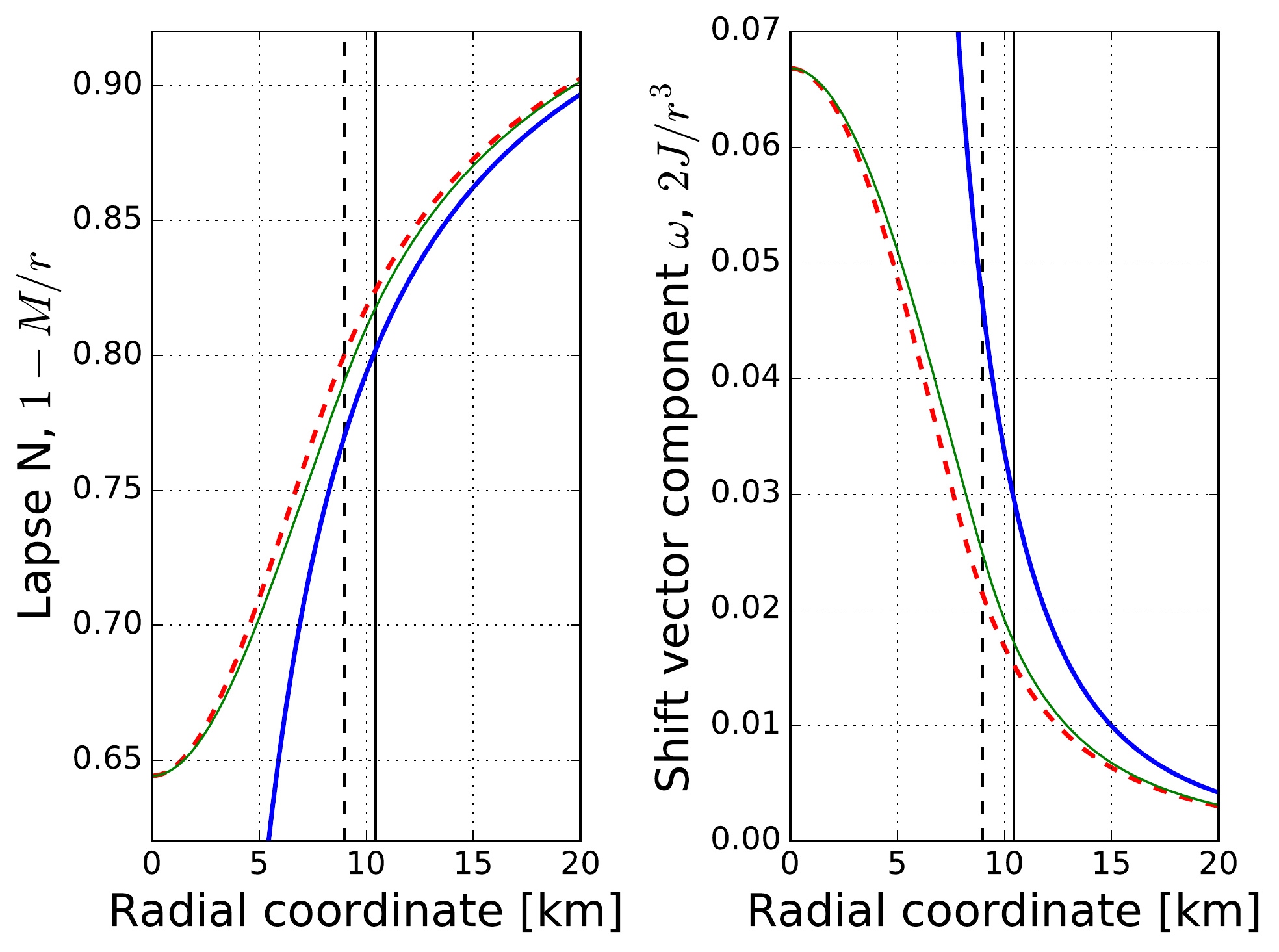}
      \caption{Left panel: Lapse function $N$ in polar (red dashed line) and equatorial
      (solid green line) direction compared with its slow-rotation approximation $1-M/r$ (solid blue).
      Right panel: Frame-dragging metric term (shift vector component $\omega = -\beta^\phi$) in polar
      (red dashed line) and equatorial (solid green line) direction compared with the slow-rotation
      approximation $2J/r^3$ (solid blue), $J$ denoting the total angular momentum of the star. Vertical lines denote the
      surface of the star: dashed for polar direction, solid for equatorial direction.}
      \label{fig:metricfunctions}
   \end{figure}

\subsection{Local spectra}
The local spectra are computed by \at for an atmosphere made of hydrogen and helium in solar abundance and an effective temperature of $T_\mathrm{eff}=10^7$~K.   
{ All models used in this paper were computed with a large number of iterations, and they achieved satisfactory convergence after 500-600 iterations depending 
on the value of surface gravity (see Appendix~\ref{appendix} for explanation)}. They are shown in Fig.~\ref{fig:localspec}. The full range of surface gravities spanned by the non-rotating and rotating configurations
has for bounds $1.19\times10^{14}$ and $1.70\times10^{14}$, in cgs units. This variation of $\approx 40\%$ of the surface gravity leads to a tiny variation of a fraction of a percent in the resulting local spectrum, which is thus scarcely sensitive to the local surface gravity.   Consequently, Fig.~\ref{fig:localspec} shows the \at spectrum at one unique value of surface gravity, equal to $g_{\rm s}=1.7\times10^{14}\,\mathrm{cm}\,\mathrm{s}^{-2}$.
Fig.~\ref{fig:localspec} also shows a color-corrected blackbody spectrum as defined by
\be
I_\nu^{\mathrm{BB\,corr}} = \frac{1}{f_\mathrm{cor}^4} B_\nu(T=f_\mathrm{cor} T_\mathrm{eff})
\ee
where $B_\nu$ is the Planck function and $f_\mathrm{cor}=1.4$ is the color-correction factor.
   \begin{figure*}
   \centering
   \includegraphics[width=\textwidth]{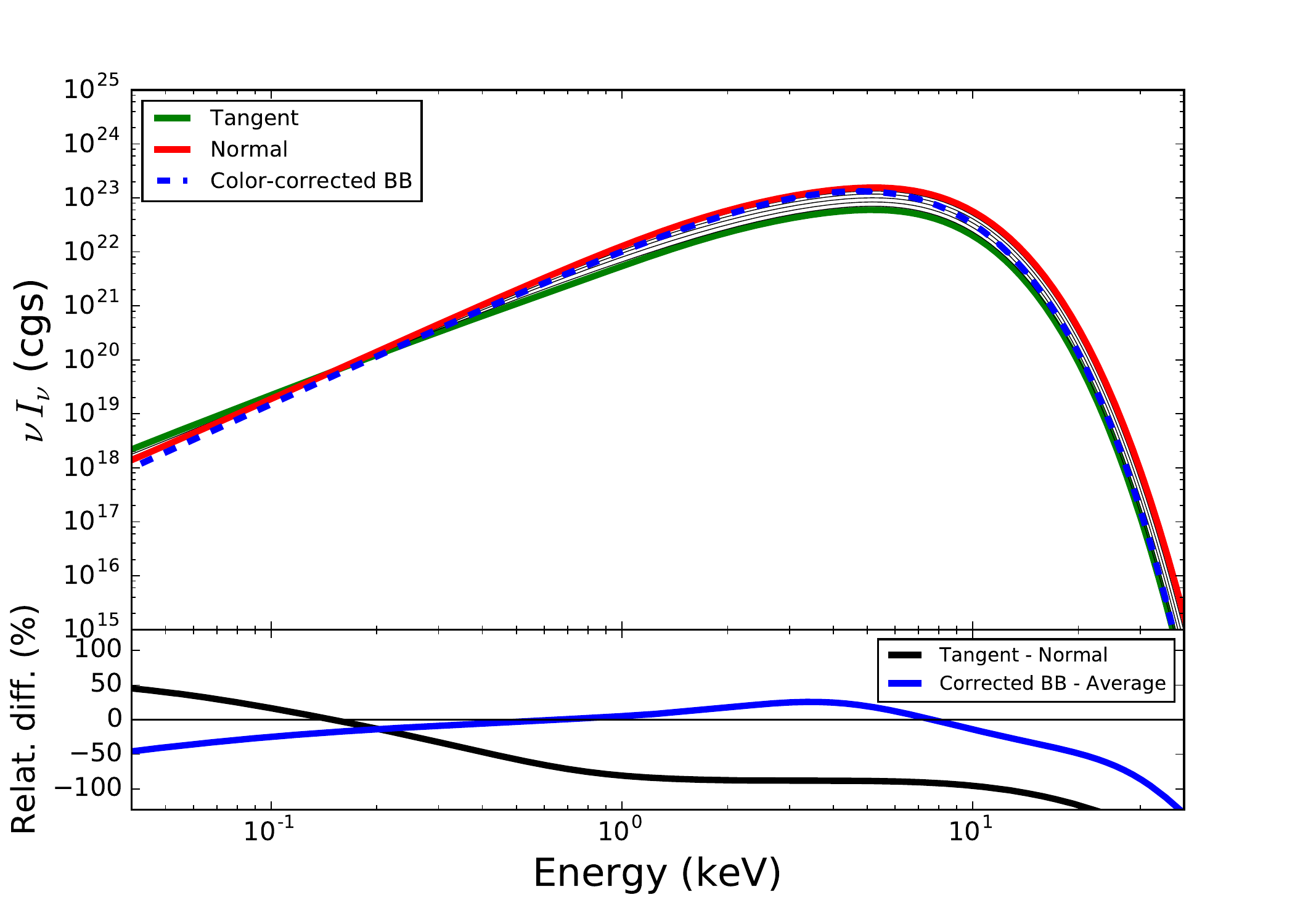}
      \caption{Local spectra emitted at the star's surface.
      \textit{Top panel}:
      the thick dashed blue curve shows a color-corrected blackbody at $T_\mathrm{eff}=10^7$~K which is also the effective temperature of the neutron star's surface used in \at. The color-correction factor is $f_\mathrm{cor}=1.4$.   All other curves are obtained by \at after solving the radiative transfer equation in the neutron star's atmosphere.   The surface gravity equals $1.7\times 10^{14}\,\mathrm{cm}\,\mathrm{s}^{-2}$ (but the effect of this parameter is very small).   Each solid thin black curve corresponds to a different emission angle $\epsilon$, with the thick red curve corresponding to emission along the local normal and the thick green curve to emission tangent to the star's surface.
      \textit{Bottom panel}:
      the percentage difference (defined as $(X_a-X_b)/(X_a+X_b)\times200$) between these two extreme angles in black (tangent minus normal), and the percentage difference between the corrected blackbody and the \at spectrum averaged over emission angle in blue.}
      \label{fig:localspec}
   \end{figure*}

Fig.~\ref{fig:localspec} also represents in its lower panel comparisons between various spectra.
It first compares the corrected blackbody spectrum to the \at spectrum averaged over the cosine of the emission angle $\mu = \cos \epsilon$.
The angle-averaged specific intensity, at frequency $\nu$ and latitude $\theta$, is 
{equal to twice the usual mean intensity}:
\be
I_\nu^{\mathrm{angle-averaged}}(\nu,\theta) =2 J_{\nu}(0)= \int_\mu I_\nu(\nu,\mu,\theta)\, \dd \mu.
\ee
The averaged \at spectrum
is overall close to the corrected blackbody, but a closer look reveals significant intensity differences at the level of $\approx 50\%$ for both the low and high energy spectral wings.
Thus, considering a realistic spectrum at the neutron star's surface not only changes the directionality of the emitted radiation, but also the shape of the spectrum.
The lower panel of Fig.~\ref{fig:localspec} also compares the tangentially- and normally-emitted local spectra.
This shows a difference with direction of order $10$s of percent, up to a factor $2$ in a large
fraction of the band. The \at spectrum is thus very strongly directional.

Effects of limb darkening, which are {typical for stars seen in the optical band}, are caused by
lowering of local gas temperature towards the interstellar space. The effect can be
mathematically derived from the formal solution of the radiative transfer equation~\citep{mihalas78}.
Most stars, including neutron stars, exhibit an inversion of
temperature which starts to rise towards the exterior. The best example is the Sun or
other late type stars which develop chromospheres with temperature inversion.
As for neutron stars, the inversion of temperature in the outermost layers of the atmosphere
is caused by the Compton scattering of radiation emitted from deeper photospheres.
For some photon energy ranges, inversion of temperature manifests itself as
limb brightening, clearly seen in our intensity spectrum for some ranges of photon frequency.



\subsection{Images}
This Section investigates the appearance of the star once imaged by our ray tracing code. We call here an image a map of observed specific intensity $I_\nu^\mathrm{obs}$.
In all the simulations presented below, the whole surface of the neutron star is assumed to emit a homogeneous
radiation $I_\nu^\mathrm{em}$.
Three ingredients are important to understand the following images. These are the impact of
\begin{itemize}
  \item the local value of the emission angle $\epsilon$;
  \item the relativistic beaming effect, that enhances the radiation of a source moving towards the observer;
  \item the local value of surface gravity $g_{\rm s}$.
\end{itemize}
The sketch of Fig.~\ref{fig:sketch-image} shows how these various quantities impact the image for an exactly edge-on view.    Edge-on views are ray-traced from an inclination of $i=90^\circ$, where $i$ is the angle between the axis perpendicular to the equatorial plane and the line of sight. These three quantities will respectively lead to a radial, horizontal and vertical gradient in the image.   All images in this section are computed at an observed energy of $4.1$~keV, close to the maximum of the neutron star's spectrum.
   \begin{figure}
   \center
      \includegraphics[width=\columnwidth]{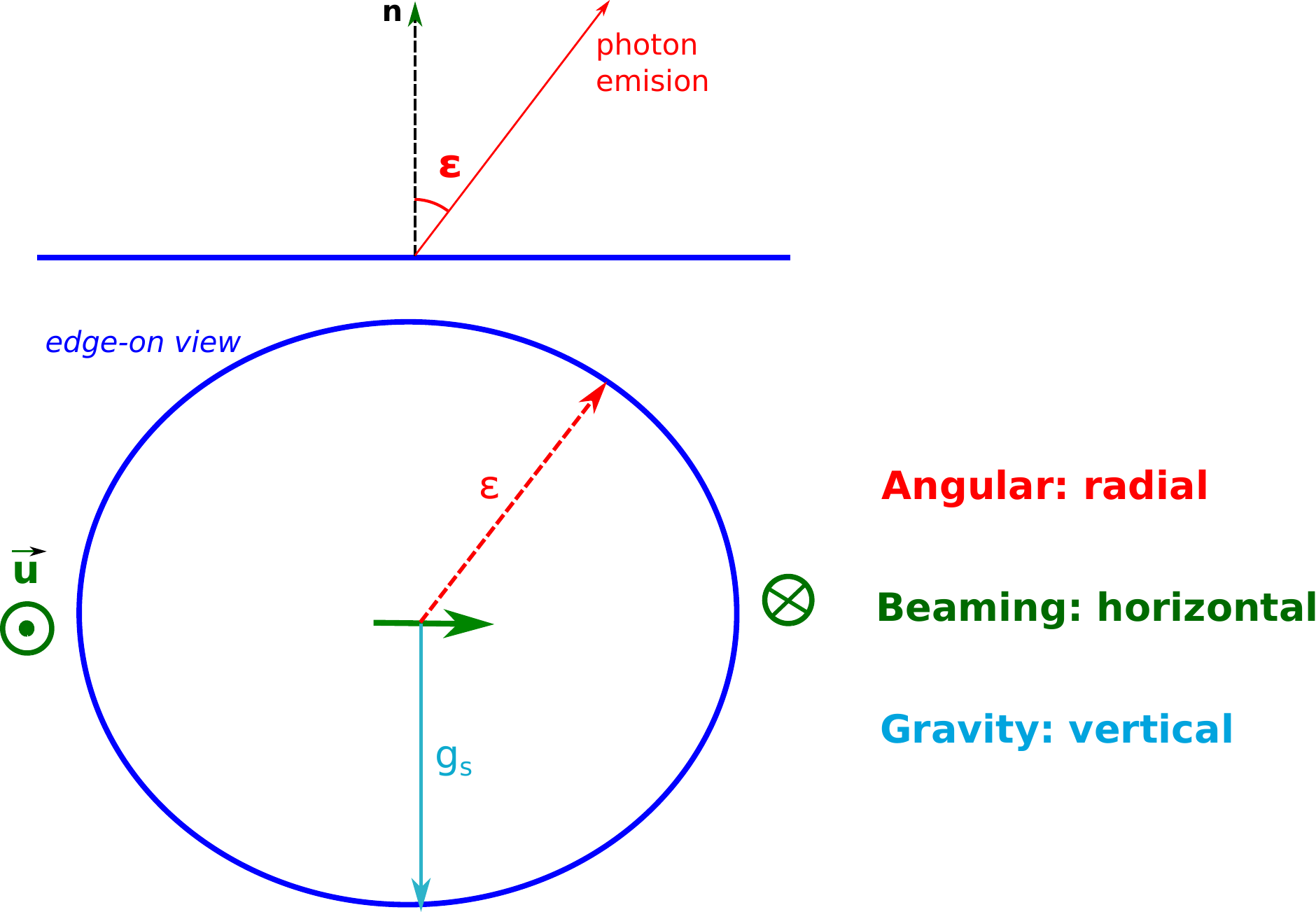}
      \caption{Sketch showing the main ingredients that explain the appearance of the star's ray-traced image.   The blue line represents the neutron star's surface.   The upper panel is a local view defining the emission angle $\epsilon$ between the local normal and the direction of photon emission.   The lower panel shows an edge-on view of a rapidly rotating neutron star.   Note that the surface is not a circle due to the high rotation.   The value of $\epsilon$ varies radially for an exactly edge-on view. {Its direction of increase is along the long dashed red arrow.}
      The 4-vector of the neutron star's surface velocity is labeled $\mathbf{u}$.   Its spatial
      part $\vec{u}$ points towards the observer on the left side, lies in the plane of the paper at the center, and points away from the observer on the right side.   The corresponding beaming effect will enhance the radiation on the left side and weaken the radiation on the right side.   The beaming effect thus shows a horizontal gradient.
The surface gravity $g_{\rm s}$ is minimum at the center of the picture (the starting point of the cyan arrow), as this point is the projection of some point in the
equator of the star where the radius is maximum. On the other hand, $g_{\rm s}$ is maximum at the end point of the cyan arrow, which
corresponds to a pole of the star, where the radius is minimum. {Thus, $g_{\rm s}$ increases from the center of the image
towards the border of the star, along the cyan vertical arrow.}
Surface gravity has constant values on horizontals in this picture, and thus shows a vertical gradient.}
      \label{fig:sketch-image}
   \end{figure}

Fig.~\ref{fig:image-i1-i90} shows the face-on and edge-on views of the non-rotating neutron star, together with the face-on view of the fast-rotating ($716$~Hz) star.   We note that we call face-on view an image obtained with an inclination $i=1^\circ$.   Our ray-tracing code uses spherical coordinates so that the $i=0^\circ$ axis is singular. We also note that the computation time of one such image (or spectrum) with resolution $30\times30$ pixels on a standard laptop takes $\approx 5$~min, allowing to compute rather easily a large number of such images/spectra.
   \begin{figure*}
      \includegraphics[width=\textwidth]{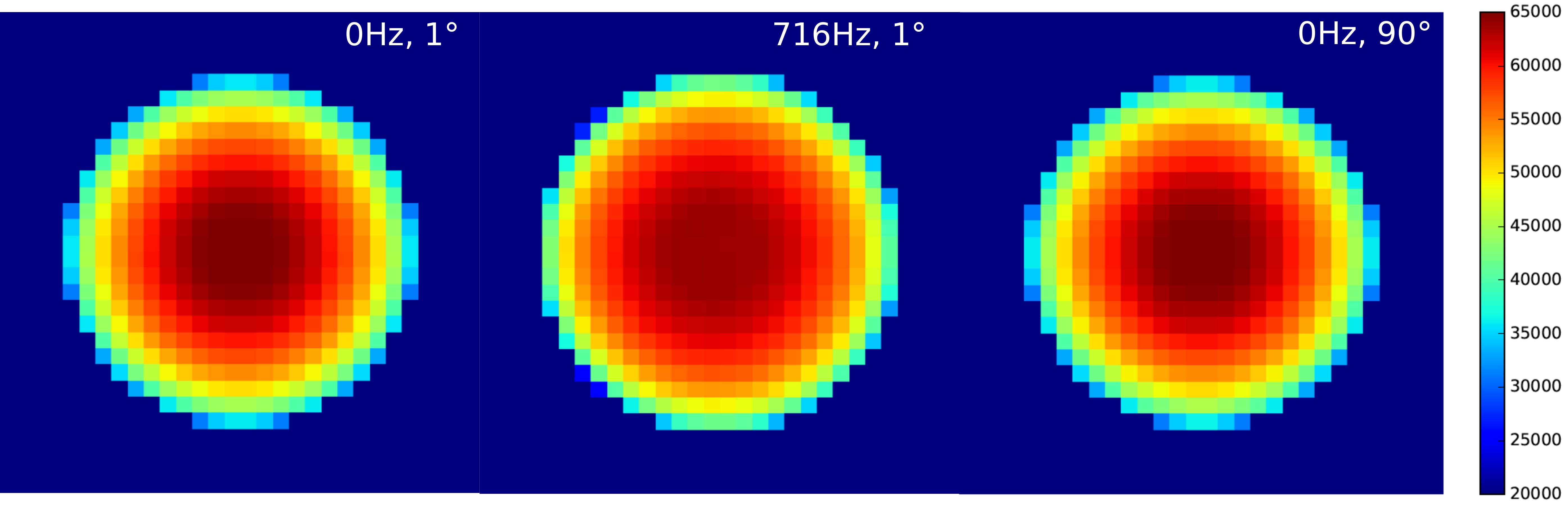}
      \caption{Images of the neutron star at observed energy $4.1$~keV with $(\Omega=0,i=1^\circ)$ for the left panel, $(\Omega=716~\mathrm{Hz},i=1^\circ)$ for the central panel, and $(\Omega=0,i=90^\circ)$ for the right panel.   The star is emitting the directional spectrum as computed by the \at code in all cases.   The mainly radial gradient of intensity is due to the radial variation of the emission angle $\epsilon$.   The color bar is common to the three panels and shows the numerical value of the observed $I_\nu$ in cgs units.}
      \label{fig:image-i1-i90}
   \end{figure*}
Obviously, the face-on and edge-on views of the non-rotating star (left and right panels) are extremely similar.   When comparing pixel by pixel, some non-negligible differences appear that can be as high as $\approx 1\%$ for a handful of pixels.   This is due to the limited precision with which the \gy code is able to find the neutron star's surface, in particular for tangential approach (remember that geodesics are integrated backwards in time, so the photon approaches the neutron star).   The code precision parameters can be adjusted to allow obtaining an arbitrarily precise result, but at the expense of computing time.   Given that the total flux, summed over all pixels, is the real quantity of interest, we are not interested in getting extremely precise values for each individual pixel.   We have checked that the flux difference between the edge-on and face-on views of the non-rotating star is of $0.05\%$, which is far better than needed to fit observations.   The mainly radial gradient of the intensity in these images is due to the radial variation of the emission angle $\epsilon$, as explained in Fig.~\ref{fig:sketch-image}.   We have checked that the image obtained when the local spectrum is averaged over emission angle shows an exactly constant value to within code precision.   The central panel of Fig.~\ref{fig:image-i1-i90} is the face-on view of the fast-rotating star.   It is very similar to the non-rotating neutron star image, which is natural as the main effect that differs when the star is rotating is the relativistic beaming effect, which is absent for a face-on view.   The pixel values are still slightly different as compared to the non-rotating case because the metric felt by the ray-traced photons is different even for a face-on view, and the surface gravity is not exactly the same in both cases.

Fig.~\ref{fig:image-rot-i90} shows edge-on images of the fast-rotating neutron star.   We have considered three different emission processes at the star's surface:   either a simple color-corrected blackbody at $T=10^7$~K, with a color-correction factor of $f_\mathrm{cor}=1.4$, or the local spectrum as computed by the \at code, averaged or not over the emission angle $\epsilon$.
   \begin{figure*}
      \includegraphics[width=\textwidth]{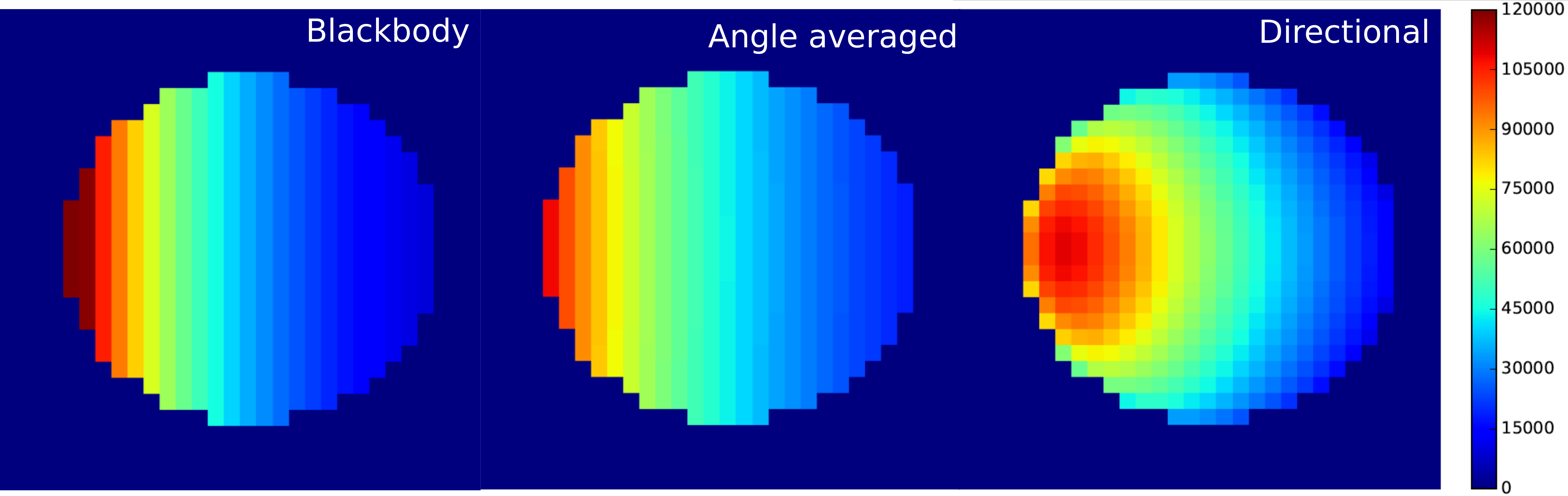}
      \caption{Edge-on images of the fast-rotating ($716~$Hz) star.    On the left panel, the star emits a simple color-corrected blackbody at $T=10^7$~K.   On the central panel, the star emits the spectrum computed by the \at code, averaged over the emission angle $\epsilon$.   The right panel shows the most realistic image with the directional spectrum as computed by \at taken into account.   The two first panels show a clear dominating horizontal gradient that is due to relativistic beaming.   The right panel superimposes to this effect       the radial gradient of the emission angle $\epsilon$, leading to a more complex appearance.   The color bar is common to the three panels and shows the numerical value of the observed $I_\nu$ in cgs units.   Note that the star's surface is not circular due to the high rotation.}
      \label{fig:image-rot-i90}
   \end{figure*}
The left and central panels are very similar.   They show a mainly horizontal gradient that is due to the relativistic beaming effect (the approaching side being on the left).   The blackbody case shows no vertical gradient, to within the code precision.   The averaged \at spectrum case shows a very small vertical gradient (not visible on Fig.~\ref{fig:image-rot-i90}, the horizontal gradient largely dominates), that is due to the slight variation of the local spectrum with surface gravity.   The right panel shows the combination of two effects:  the horizontal gradient due to relativistic beaming, and the radial gradient due to the variation of the emission angle $\epsilon$.   The superimposition of these two effects leads to a more complex image.

\subsection{Ray-traced spectra}
This Section investigates the end product of our numerical pipeline, i.e. the ray-traced neutron star spectra.   All the spectra in this Section are computed with the same $30\times30$ pixels resolution as the images presented in the previous Section. We checked by comparing to a $100\times100$ resolution for one particular case that this resolution
leads to an error smaller than $0.7\%$ over the full energy range.

\subsubsection{Impact of the emission process}
Let us first focus on the impact of the local radiation process at the neutron star's surface on the end-product, ray-traced observed spectrum.   Fig.~\ref{fig:spec_emisprocess} shows the ray-traced spectra of the non-rotating or fast-rotating neutron star for three different kinds of emissions at the star's surface:   either a color-corrected blackbody at the effective temperature of $T=10^7$~K, or the \at spectrum averaged over emission angle, or the directional \at spectrum.

   \begin{figure*}
    \centering
            {\includegraphics[width=15cm]{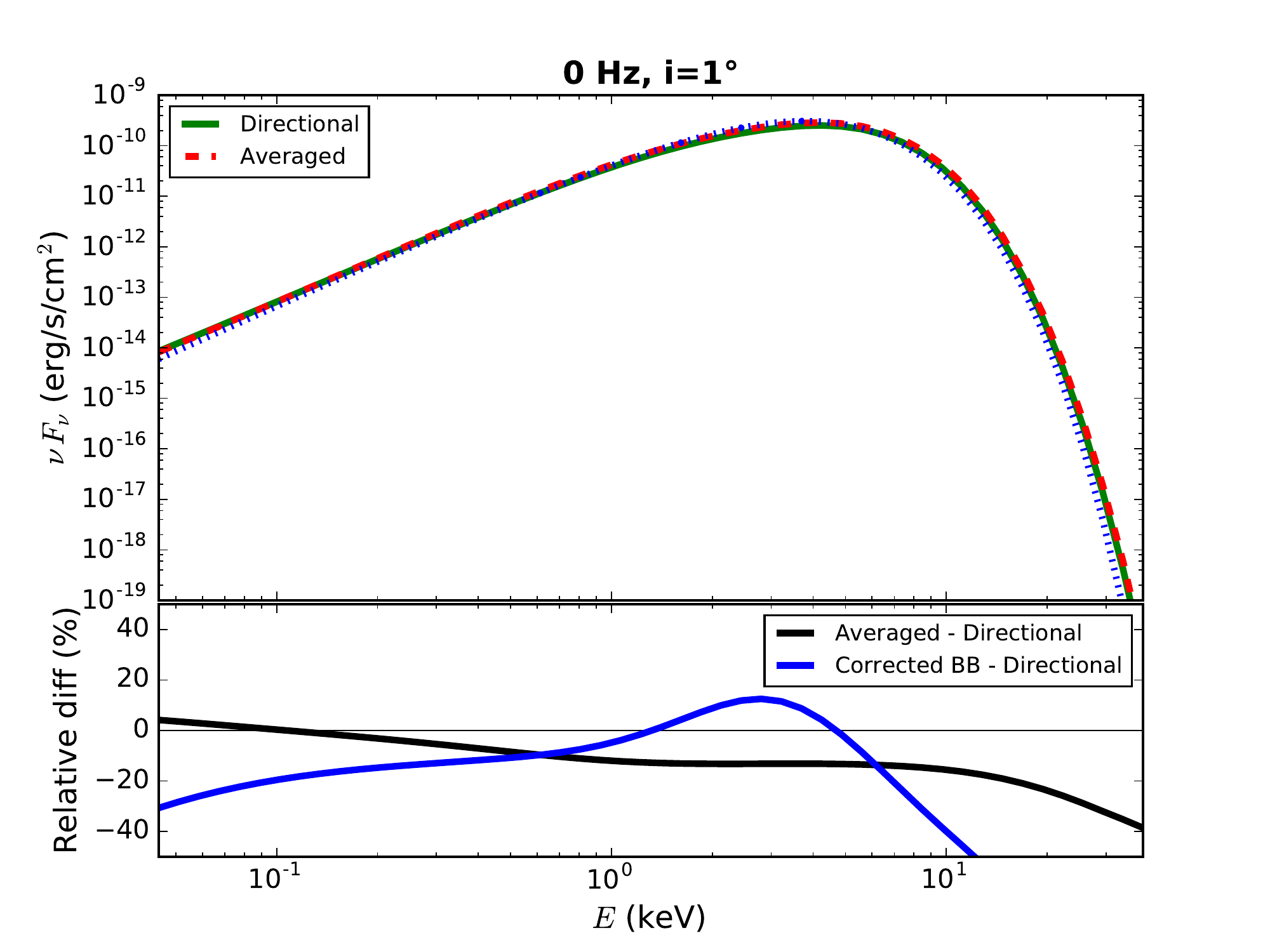}}
            {\includegraphics[width=15cm]{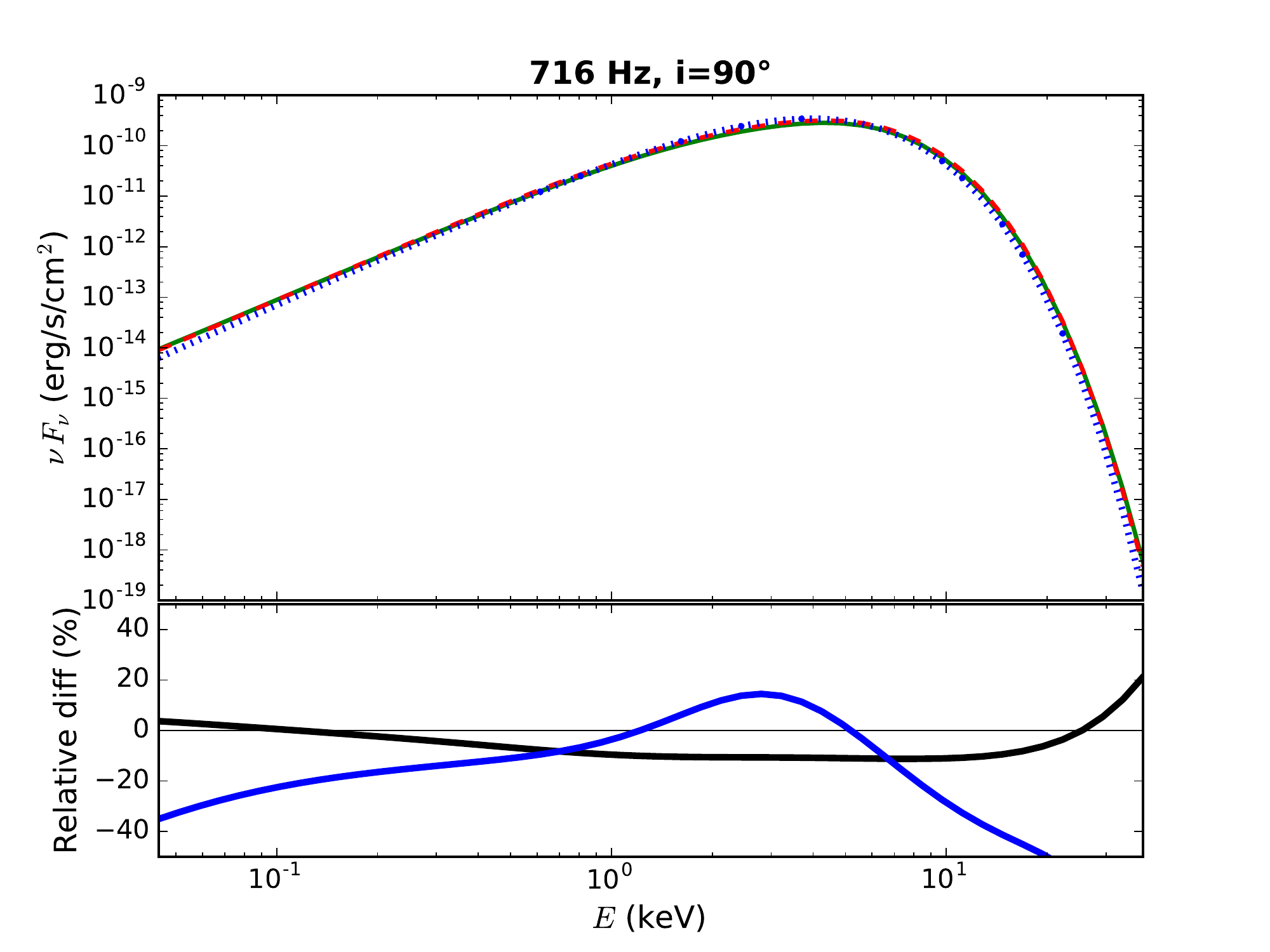}}
      \caption{\textbf{Effect of the emission process on the observed spectra:}
      \textit{Top panel:}
      ray-traced spectra (and relative differences) of a neutron star with zero rotation seen face on.
      \textit{Lower panel: }
      same for a neutron star with $716$~Hz rotation seen edge on.   Three different radiation processes are considered at the star's surface: color-corrected blackbody at $T=10^7$~K (dotted blue),  \at spectrum averaged over emission angle (solid green), or \at directional spectrum (dashed red).    The lower sub-panel of each plot shows the relative difference in percentage between the averaged and directional \at spectra (solid black), and between the corrected blackbody and the directional \at spectrum (solid blue).}
      \label{fig:spec_emisprocess}
   \end{figure*}

This Figure shows that the difference between the ray-traced spectra of a neutron star emitting a color-corrected blackbody or a realistic spectrum as computed by \at is of the same order as what was obtained in the local context of Fig.~\ref{fig:localspec}. Thus, taking a realistic emitted spectrum into account changes the observed spectrum by typically $\approx 20\%$ over most of the energy band.
This difference is related to the fact that the locally emitted spectrum is strongly directional. This directional dependence cannot be simplified,
particularly when the star rotates fast and the photons undergo highly non-trivial lensing effects that impact their emission angle $\epsilon$,
and thus the specific intensity they carry. We note that, as illustrated in Fig.~\ref{fig:localspec}, even the angle-averaged local spectrum is still very different
from a corrected blackbody, so that even if one replaces the directional emitted spectrum by its angle average (which would be wrong, because
of the non-trivial lensing effects highlighted before), the resulting ray-traced spectrum would still be incorrect.
This problem is analogous to the computation of the spectra reflected by black holes' accretion disks.
Ray-tracing techniques
must be used to compute the reflected spectrum in order to take properly into account the strong directional dependence of the local spectrum~\citep{garcia14,vincent16}.





Figure~\ref{fig:spec_emisprocess} also allows to compare the end-product ray-traced spectrum when averaging the \at local spectrum over emission angle as compared to the directional case.   This is particularly interesting as a test of our pipeline, because the result can be predicted to some extent.   Let us first focus on the $(\Omega=0, i=1^\circ)$ case (that is very similar to the $(\Omega=716~\mathrm{Hz}, i=1^\circ)$ case, which is thus not shown here).
The relative difference between the ray-traced averaged and directional spectra is very similar to the relative difference between the local tangent and normal spectra of Fig.~\ref{fig:localspec} (solid black, lower panel):
\be
\underbrace{\mathrm{Averaged - Directional}}_{\mathrm{in \;ray-traced \;spectrum}} \approx \underbrace{\mathrm{Tangent - Normal}}_{\mathrm{in \;local \;spectrum}}.
\ee
The absolute value of the relative difference is smaller in the ray-traced case, but the trend is exactly the same, as if the directional ray-traced spectrum would privilege normal emission.   It is indeed so, as illustrated on Fig.~\ref{fig:area}: the projection effect suffered by the star when it is imaged on a flat screen at the observer's  position
   \begin{figure}
            {\includegraphics[width=\columnwidth]{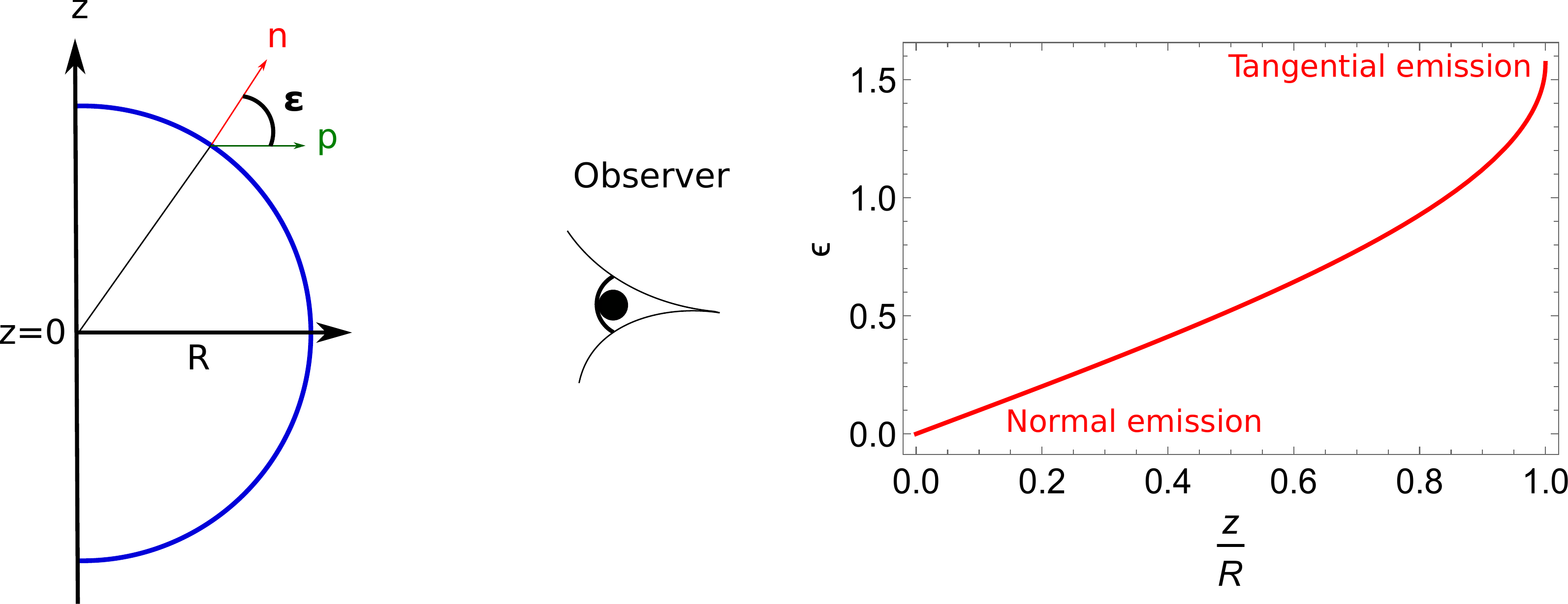}}
      \caption{On the left, the star's surface is shown in blue.   The $z$ axis is the axis of rotation, $\mathbf{n}$ is the local normal at some point of the surface, $\mathbf{p}$ is the direction of emission towards the observer, $\epsilon$ is the emission angle, and $R$ the neutron star radius.   The plot on the right shows the evolution of $\epsilon = \mathrm{arcsin} (z/R)$ with $z/R$.   The emission is normal when $\epsilon=0$ and tangential when $\epsilon=\pi/2$. Because of the projection effect when imaging the neutron star on a screen at the observer's location, more weight will be given to normal emission than to tangential emission.  {Indeed, a majority of values of $z/R$ give rise to values of $\epsilon$
      that are closer to $0$ than to $\pi/2$, as is clear from the right panel.} }
      \label{fig:area}
   \end{figure}
leads to giving more weight to normal emission than to tangential emission.   On the contrary, when considering an averaged spectrum, all values of $\epsilon$ are considered with the same weight.   The trend of the top panel of Fig.~\ref{fig:spec_emisprocess} thus makes sense.   Let us now focus on the lower panel of this Figure, i.e. the $(\Omega=716~\mathrm{Hz},i=90^\circ)$ case.   The relative difference between the averaged and directional ray-traced spectra is similar to the top panel on the low-energy side, but then gets inverted at high energies.   This is linked to the relativistic beaming effect, which is the main effect shaping the edge-on spectra as illustrated in Fig.~\ref{fig:image-rot-i90}.   However, the effect of beaming strongly depends on the spectrum's slope.    Let us illustrate this by focusing on the approaching side of the star (the left part of the image in Fig.~\ref{fig:image-rot-i90}).   As the emitter travels towards the observer, the Doppler effect translates in the observed photon energy being higher than the emitted one.   Thus, the Doppler effect pushes the emitted intensity towards smaller values when the spectrum increases with energy, and towards higher values when the spectrum decreases with energy (see Fig.~\ref{fig:beamingspec}).
   \begin{figure}
   \center
            {\includegraphics[width=\columnwidth]{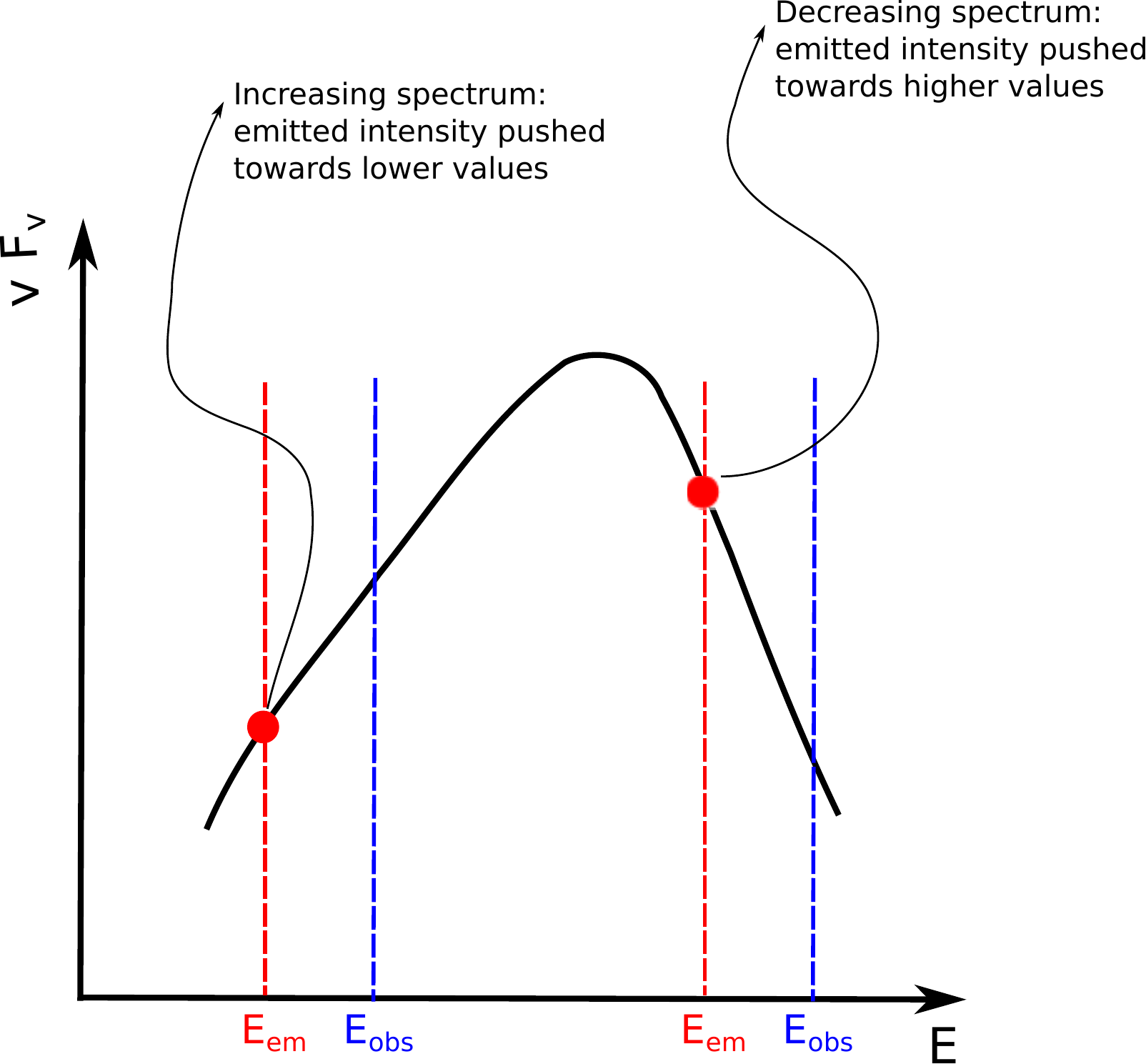}}
      \caption{Illustration of the combination of Doppler and beaming effect depending on the spectral trend, on the approaching side of a neutron star observed edge-on.   See text for details.}
      \label{fig:beamingspec}
   \end{figure}
On the approaching side, the beaming effect has always the same effect, whatever the spectrum shape, it enhances the radiation. Consequently, Doppler effect tends to compensate the beaming effect on the increasing side of the  spectrum (i.e. low-energy side), and to boost the beaming effect on the decreasing side of the spectrum (i.e. high-energy side).   This allows to understand the trend of the relative difference between the averaged and directional spectra of the lower panel of Fig.~\ref{fig:spec_emisprocess}.   On the low-energy side, Doppler effect counterbalances the beaming effect so that things look very much like the face-on case which is not affected by beaming. On the contrary, on the high-energy side, Doppler effect increases the effect of beaming, strongly  concentrating the radiation in the left-most part of the image.   This part of the image is associated to mostly tangential emission as it lies far away from the center of the neutron star on the observer's screen.   Thus, the directional
ray-traced spectrum selects tangential emission, the opposite as to what was the case in the face-on scenario  discussed above.   This explains the opposite trend of the relative difference plot in the lower panel of Fig.~\ref{fig:spec_emisprocess}, as compared to the top one.   Thus, all trends depicted on Fig.~\ref{fig:spec_emisprocess} can be explained.   This is a strong argument in favor of the correctness of our pipeline.

\subsubsection{Impact of the angular velocity}
Figure~\ref{fig:spectra_rot} illustrates the effect of the star's rotation on the observed ray-traced spectrum, when the neutron star emits either a color-corrected blackbody or the directional \at spectrum.
   \begin{figure*}
            {\includegraphics[width=9.5cm]{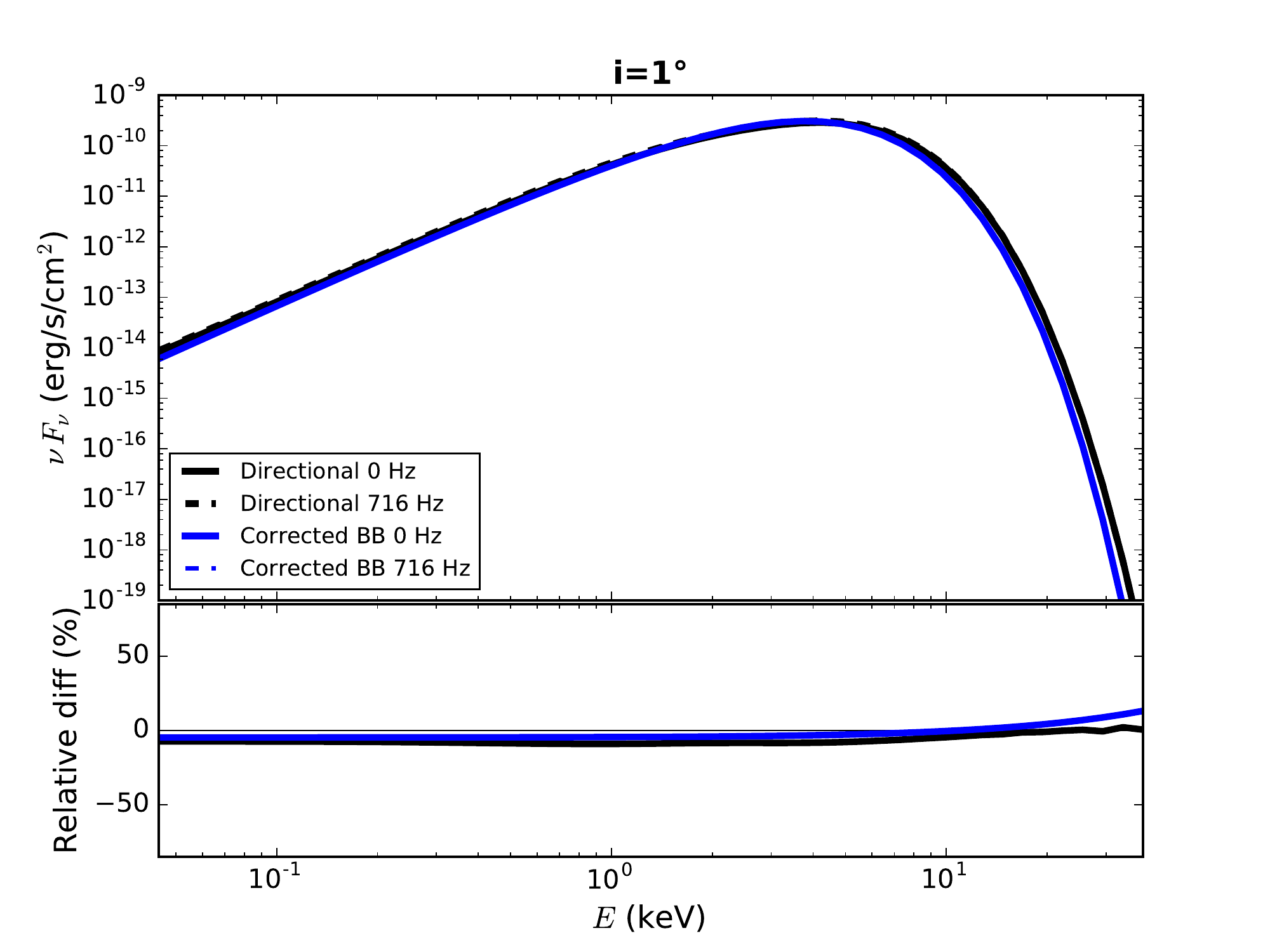}}
            {\includegraphics[width=9.5cm]{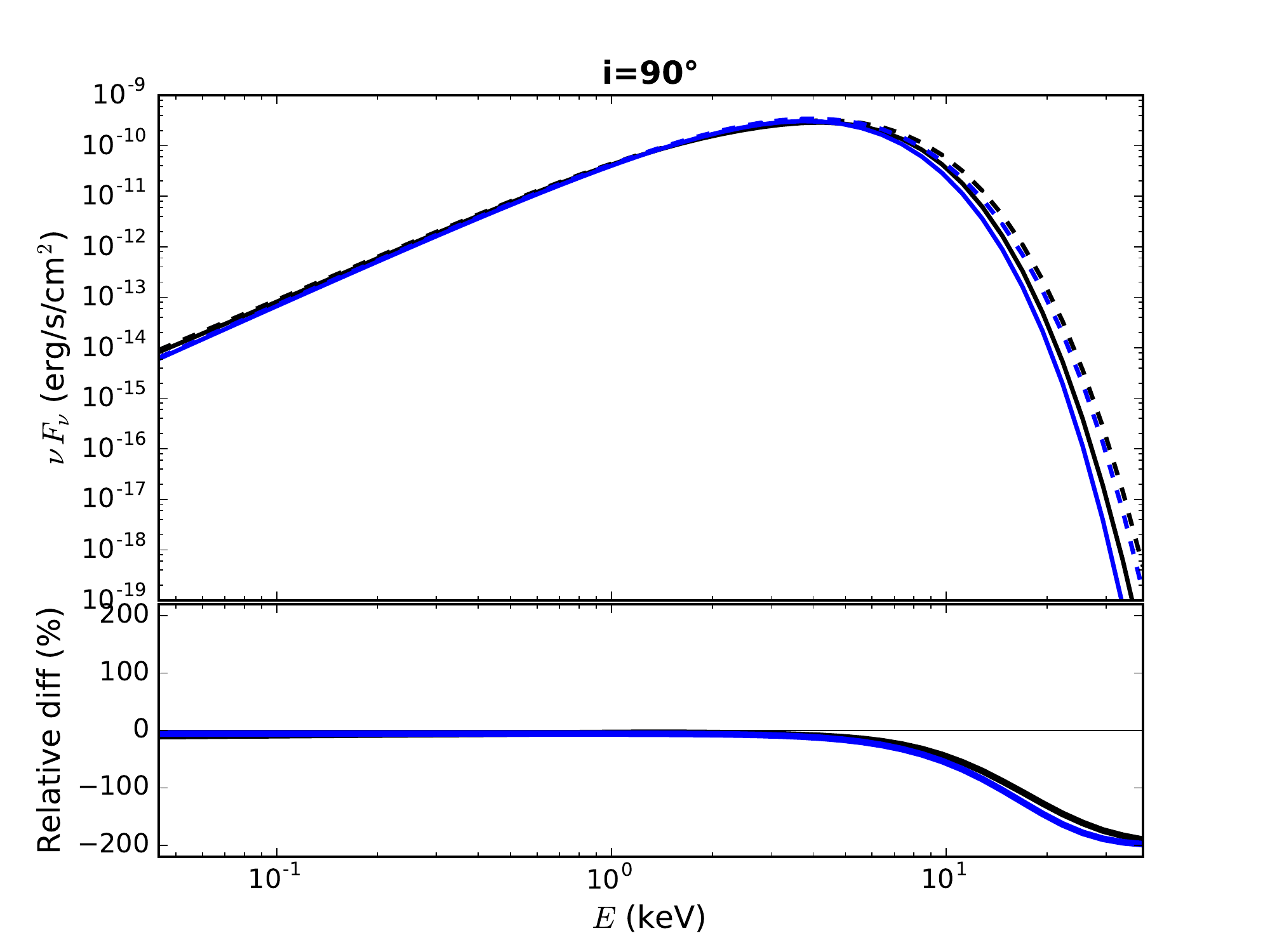}}
      \caption{\textbf{Effect of rotation on the observed spectrum:}  ray-traced spectra of a neutron
      star seen face on (left) or edge on (right).   The star's surface emits color-corrected blackbody (blue) or the directional \at spectrum (black) and rotates at $0$~Hz (solid) or $716$~Hz (dashed). The lower panels show the relative difference between the two blackbody and directional spectra in blue and black respectively.   Note the change of the vertical axis scale between the two lower panels.}
         \label{fig:spectra_rot}
   \end{figure*}
It shows that the rotation has a small impact when the star is seen face on, which is not surprising after what has been discussed on Fig.~\ref{fig:image-i1-i90}.   The difference is here purely due to the change of the spacetime metric with the star's rotation.   On the contrary, the effect of rotation is huge
at high energy when the star is seen edge on, with a difference reaching  factors of a few. However, this effect is still small, even for the edge-on case, on the low-energy side. This trend is explained in the same way as above, by the coupled influence of the Doppler and beaming effects.   Fig.~\ref{fig:spectra_rot} also shows that the effect of rotation is very similar for the two different emission processes, blackbody and directional \at spectrum.

{The bolometric flux relative difference between the $0$~Hz and $716$~Hz cases, for edge-on view, is of $13.7\%$ for the blackbody
case, and $11.2\%$ for the \at case. These numbers are difficult to directly compare to the results of~\citet{baubock15}
given that the neutron star's parameters are not the same. However, we note that the bolometric relative difference
reported by these authors between a $15$~km neutron star rotating at $\approx 700$~Hz and its non-rotating counterpart, seen edge-on, is of $\approx 15\%$ according
to their Fig.~4, lower panel.}




\section{Conclusions}
\label{sec:conc}

We have presented a new numerical pipeline that intends to compute very
accurate X-ray bursting spectra of neutron stars, for arbitrary masses and rotation.
It is the only tool in the literature, so far, that allows to compute neutron-star spectra
considering realistic model atmospheres together with all general-relativistic
effects on the star's shape and photon propagation. Our pipeline is the concatenation
of the \lonr code together with the \at and \gy codes. At the present time, only \lonr and \gy are
fully open-source, but we consider making \at open-source in the close future,
so that the pipeline can be freely used by the community.

This article first demonstrates the validity of our pipeline by investigating in detail
its outputs and comparing them when possible to the literature. It also highlights the importance
of considering both a precise directional model atmosphere and ray tracing in order
to obtain accurate predictions.

Our future work will be dedicated to broadening the astrophysical impacts
of this new pipeline, making it a testbed to investigate
the validity of the various simplifying
assumptions, either on the emitted spectrum or on the treatment of strong gravity,
that other codes consider. The output of the pipeline will be implemented in {\sc xspec} by creating
fits tables of bursting neutron star spectra that might then be easily used for fitting data.
Finally, we identify potentially interesting research directions that can be only undertaken with
such a fully relativistic pipeline, namely
detailed investigation of the impact of the dense matter EoS on the observables.
We will also consider "extreme" neutron stars, both very massive and fast-rotating,
in order to predict the observable specific features of these very special sources.

\acknowledgments
{\small
FHV thanks J\'er\^ome Novak and Micaela Oertel for interesting discussions.
AR, JM and BB were supported by Polish National Science
Center grants No. 2015/17/B/ST9/03422, 2015/18/M/ST9/00541.
MB, MF, PH, and LZ acknowledge support from 2013/11/B/ST9/04528.}

\software{
\lo~\citep{lorenelib}, 
\at~\citep{madej04,majczyna05}, 
\gy~\citep{vincent11}.} 

\appendix

\section{Validity of \at code}
\label{appendix}
Numerical solution of {the model atmosphere} problem is very time consuming.
{We assume, that each model fulfills the conditions of radiative and
hydrostatic equilibrium, which determines the evolution of temperature and gas pressure  
in the atmosphere. These two conditions are numerically achieved by means of iterations,
with gas pressure iterations being nested in temperature iterations.} 

{\at uses very efficient and fast algorithms to ensure hydrostatic
equilibrium. However, temperature iteration steps are very slow and must be
repeated hundreds of times in the most complicated models. In a typical computation,
each step requires solving 1025 (i.e. the number of
frequencies) integro-differential equations of transfer with Compton redistribution function 
at all optical depth levels (typically 96 levels). }

The \at code is constantly upgraded to increase its accuracy.
The most difficult case is the pure hydrogen atmosphere, when matter is completely ionized and Compton scattering is the dominant process of opacity. Furthermore, the convergence problem appears when the surface gravity  is close to the Eddington limit, above which the atmosphere becomes unstable.

Recently \citet{suleimanov12} pointed out that our code \at is not correct on the basis of a single result, which was exchanged with those authors by us in private communication.
This result was obtained for a pure hydrogen atmosphere, with parameters
close to the Eddington limit {$T_{\rm eff} =1.80\times 10^7$ K and $\log g =14.0$  (cgs units)}. Even if flux and temperature corrections were iterated with the accuracy of  0.1\%  the final spectral shape was not converged yet (after 62 temperature iterations, which we normally have used for testing purposes).  In Fig.~\ref{fig:valery} we show also subsequent iterations, 264 and 1954, of the same model atmosphere computed with the same version of \at code as quoted by \citet{suleimanov12}.
The above figure shows significant evolution of the model spectrum reaching the
convergence limit of about 2000 iterations. {For models with higher surface gravity 
(i.e. far from Eddington limit) usually less than 600 iterations are sufficient to achieve 
satisfactory convergence. The remaining difference with respect to \citet{suleimanov12} is displayed  at the bottom panel of Fig.~\ref{fig:valery}. }

{Therefore, the fully iterated sample \at model is also in agreement with 
\citet{boutloukos2010} and \citet{miller2011}, who concluded that the
brightest RXTE spectra of X-ray bursters 4U 1820-30 and GX 17+2 are best fitted
by Bose-Einstein spectra possibly with nonzero chemical potential.}

We note, that  the spectral shape computed with  \at code and discussed  by \citet{ suleimanov12} and by \citet{medin16} was neither published nor endorsed for publication by our group in any of our earlier papers. {The differences between models computed
by different groups may be caused by the use of different algorithms or computational
schemes since the problem is very complex and requires huge CPU time. 
A more detailed comparison of \at models and those of \citet{ suleimanov12}
will be presented in a forthcoming paper, Madej et al. (2018, in prep.).}

 \begin{figure}
   \center
      \includegraphics[width=15cm]{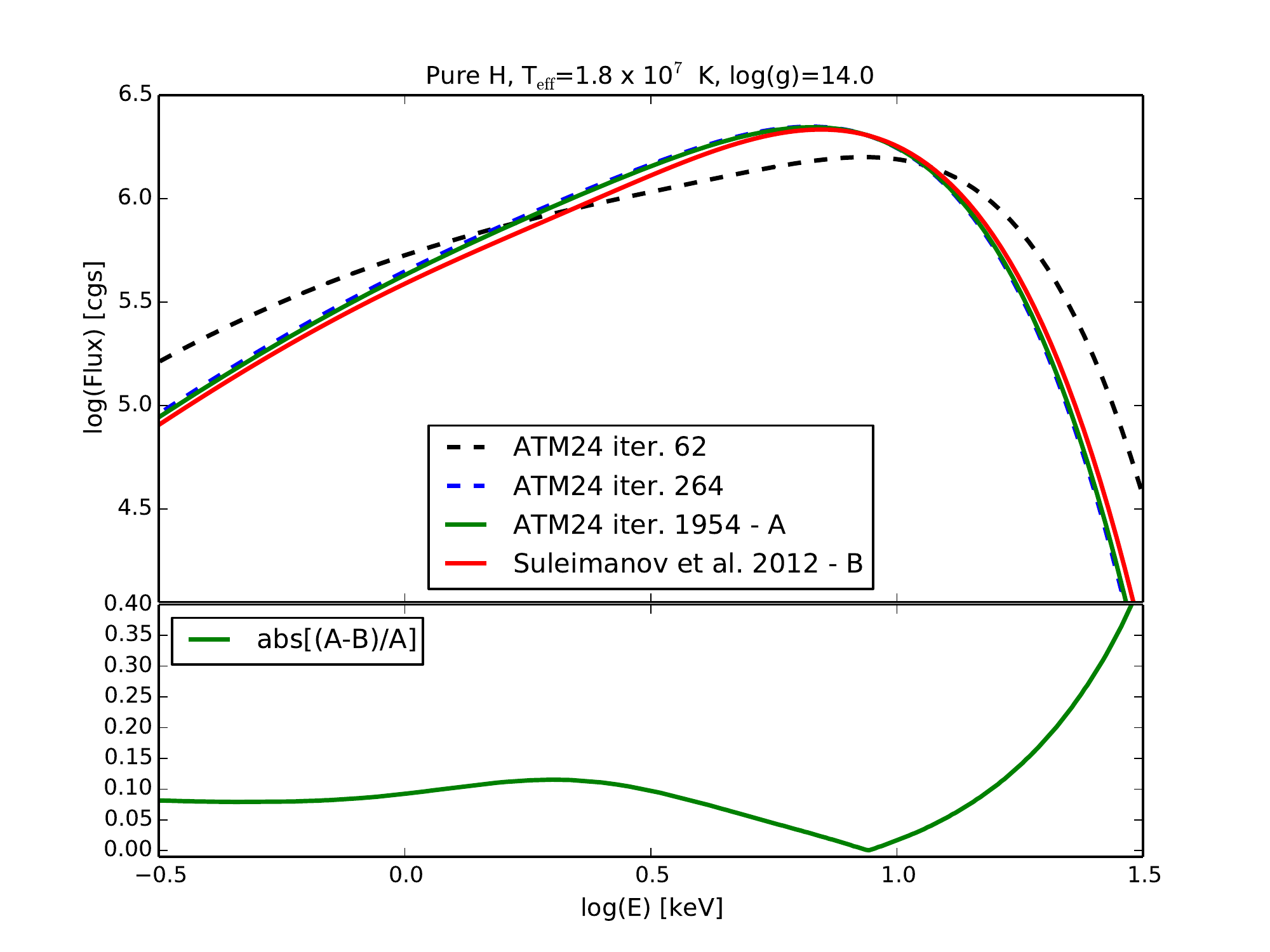}
      \caption{The convergence of \at code for pure hydrogen atmosphere. The dashed black line shows the spectrum after 62 iterations, which is not fully converged, and was
      used in the comparison of Fig.~C.1 of~\cite{suleimanov12}. Increasing the number of iterations
      to 264 and 1954 respectively gives the dashed blue and solid green lines. These spectra show perfect convergence of spectral shape {and are compared} to the spectrum computed by~\citet{suleimanov12} numerical code (solid red line).
      {The comparison shows a very good agreement at the few percent level
      around the spectrum maximum.}
      }
      \label{fig:valery}
   \end{figure}

\bibliographystyle{aasjournal}
\bibliography{references}

\end{document}